\def\build#1_#2^#3{\mathrel{
\mathop{\kern0pt#1}\limits_{#2}^{#3}}}
\def\ga{\mathrel{\mathchoice {\vcenter{\offinterlineskip\halign{\hfil
$\displaystyle##$\hfil\cr>\cr\sim\cr}}}
{\vcenter{\offinterlineskip\halign{\hfil$\textstyle##$\hfil\cr>\cr\sim\cr}}}
{\vcenter{\offinterlineskip\halign{\hfil$\scriptstyle##$\hfil\cr>\cr\sim\cr}}}
{\vcenter{\offinterlineskip\halign{\hfil$\scriptscriptstyle##$\hfil
\cr>\cr\sim\cr}}}}}
\def\la{\mathrel{\mathchoice {\vcenter{\offinterlineskip\halign{\hfil
$\displaystyle##$\hfil\cr<\cr\sim\cr}}}
{\vcenter{\offinterlineskip\halign{\hfil$\textstyle##$\hfil\cr<\cr\sim\cr}}}
{\vcenter{\offinterlineskip\halign{\hfil$\scriptstyle##$\hfil\cr<\cr\sim\cr}}}
{\vcenter{\offinterlineskip\halign{\hfil$\scriptscriptstyle##$\hfil
\cr<\cr\sim\cr}}}}}

\def\bar{\overline}
\def\ben{\begin{equation}}
\def\be#1{\begin{equation}\label{eq:#1}}
\def\ee{\end{equation}}
\def\EC#1{(\ref{eq:#1})}

\documentstyle[11pt,aaspp4,fleqn,epsf]{article}
\def\build#1_#2^#3{\mathrel{
\mathop{\kern0pt#1}\limits_{#2}^{#3}}}
\def\la{\mathrel{\mathpalette\fun <}}
\def\ga{\mathrel{\mathpalette\fun >}}
\def\fun#1#2{\lower3.6pt\vbox{\baselineskip0pt\lineskip.9pt
        \ialign{$\mathsurround=0pt#1\hfill##\hfil$\crcr#2\crcr\sim\crcr}}}

\def\bar{\overline}

\def\ben{\begin{equation}}
\def\be#1{\begin{equation}\label{eq:#1}}
\def\bea#1{\begin{eqnarray}\label{eq:#1}}
\def\ee{\end{equation}}
\def\eea{\end{eqnarray}}
\def\EC#1{(\ref{eq:#1})}

\def\vp{\varpi}
\def\kpc{{\rm \,kpc}}
\def\kms{{\rm \,km/sec}}
\def\th{\hat t}
\def\pkdk{{\cal P}_k \Delta_k}
\def\pjdj{{\cal P}_j \Delta_j}
\def\vndn{{\cal V}_n \Delta_n}
\begin{document}

\begin{titlepage}
\null\vspace{-62pt}
\vspace{1.5in}
\baselineskip 12pt
\centerline{{\Large \bf 
Searching for Machos (and other Dark Matter Candidates)}}
\medskip
\centerline{~{\Large \bf in a Simulated Galaxy}}
\vspace{0.2in}
\centerline{{\bf Lawrence M. Widrow}\footnote{E-mail address:
widrow@astro.queensu.ca}}
\vspace{0.1in}
\centerline{{\it Department of Physics}}
\centerline{{\it Queen's University, Kingston, K7L 3N6, CANADA}}
\vspace{.25in}
\centerline{and}
\vspace{.25in}
\centerline{~{\bf John Dubinski}\footnote{E-mail address:
dubinski@cita.utoronto.ca}}
\vspace{0.1in}
\centerline{{\it Canadian Institute for Theoretical Astrophysics}}
\centerline{{\it University of Toronto, Toronto, M5S 1A1, CANADA}}
\vspace{.5in}

\begin{abstract}

We conduct gravitational microlensing experiments in a galaxy taken
from a cosmological N-body simulation.  Hypothetical observers measure
the optical depth and event rate toward hypothetical LMCs and compare
their results with model predictions.  Since we control the accuracy
and sophistication of the model, we can determine how good it has to
be for statistical errors to dominate over systematic ones.  Several
thousand independent microlensing experiments are performed.  When the
``best-fit'' triaxial model for the mass distribution of the halo is
used, the agreement between the measured and predicted optical depths
is quite good: by and large the discrepancies are consistent with
statistical fluctuations.  If, on the other hand, a spherical model is
used, systematic errors dominate.

Even with our ``best-fit'' model, there are a few rare experiments
where the deviation between the measured and predicted optical depths
cannot be understood in terms of statistical fluctuations.  In these
experiments there is typically a clump of particles crossing the line
of sight to the hypothetical LMC.  These clumps can be either
gravitationally bound systems or transient phenomena in a galaxy that
is still undergoing phase mixing.  Substructure of this type, if
present in the Galactic distribution of Machos, can lead to large
systematic errors in the analysis of microlensing experiments.

We also describe how hypothetical WIMP and axion detection experiments
might be conducted in a simulated N-body galaxy.

\end{abstract}

\keywords{galaxies: halo --- galaxies: structure --- cosmology: dark
matter --- methods: numerical}

\end{titlepage}
\newpage

\section{Introduction}

Four years ago the MACHO (Alcock et al.\,1993) and EROS (Aubourg et
al.\,1993) collaborations announced candidate gravitational
microlensing events toward the Large Magellanic Cloud (LMC)
demonstrating the viability of a new and potentially powerful probe of
dark matter in the Galactic halo.  Microlensing experiments
(Paczy\'{n}ski 1986; Griest 1991) are sensitive to any object that is
smaller than its Einstein radius (in the halo, these objects are known
as Machos for massive compact halo objects) and therefore complement
direct observations, which survey the visible content of the Galaxy,
and dynamical studies, which measure the total mass density.
Unfortunately, microlensing experiments are subject to a number of
limitations which make interpretation of their results rather
difficult.  First, microlensing events are extremely rare and it is
necessary to monitor $O(10^6)$ stars in order to get just a few events
per year.  This has restricted present day searches to regions of the
sky where there are dense concentrations of stars.  Indeed the only
published microlensing events that can be associated with halo objects
have been toward the Large and Small Magellanic Clouds (Alcock et
al.\,1993, 1995, 1996, 1997b; Aubourg et al.\,1993).  Microlensing
experiments, with only one or two lines of sight, tell us little about
the structure of the halo and, more to the point, are sensitive to
potentially large systematic errors due to our incomplete knowledge of the
halo's structure.  Moreover, because of the small number of events,
there are large statistical uncertainties.  Finally, present day
experiments are unable to determine unambiguously the mass and
velocity of a given lens.  Despite these limitations the MACHO
collaboration estimates, from the two-year data set, that $\sim 50\%$
or more of the mass in the halo within $50\kpc$ is composed of $\sim
0.5\,M_\odot$ objects (Alcock et al.\,1996).  If true, these results
would have important implications for cosmology, galaxy formation, and
star formation.

The mass density in Machos is determined by comparing the observed
number of events with the number predicted for a particular model of
the Galaxy.  Potential systematic errors are estimated by seeing how
the predicted number of events varies for different ``reasonable''
Galactic models.  This strategy, however, is hindered by our limited
ability to construct and analyze the full complement of acceptable
models.  In particular, the models used to analyze MACHO's results do
not generally include triaxiality, velocity space anisotropy, and
substructure, all of which are expected in the real distribution of
Machos.

In this paper we propose an alternative approach to understanding
gravitational microlensing experiments.  Microlensing experiments are
conducted in a halo taken from a cosmological N-body simulation.
Hypothetical observers make measurements of the optical depth and
event rate and compare their results with model predictions much as
real observers would.  We can perform a large number of microlensing
experiments on a single N-body halo by simply changing the positions
of the observer and LMC.  Since we control the accuracy and
sophistication of the model, we can determine how good it has to be
for statistical errors to dominate over systematic ones.

Our intention is not to create an N-body realization of the Milky Way
Galaxy.  Indeed our galaxy bears little resemblance to the Milky Way.
Moreover, the simulation assumes a cold dark matter universe which may
not be appropriate if a significant fraction of the dark matter is
composed of Machos.  However the distribution of particles in the
simulated galaxy does exhibit the general characteristics (e.g.,
triaxiality, substructure) that one expects in a realistic
distribution of Machos.  We can therefore test how well theory agrees
with observation when these features are not fully taken into account.

We find that a simple triaxial model, with the axial ratios and
density profile determined from the simulation, provides excellent
agreement between measured and predicted optical depths.  In all but a
few rare experiments the discrepancies are consistent with the
statistical fluctuations that one would expect had the particles been
chosen at random from the model distribution function.  If instead an
axisymmetric spheroidal model is used, systematic errors become
important.  And if a spherical model is used, systematic errors
dominate.

Microlensing experiments are also sensitive to the velocity space
distribution of the Machos.  If the velocities assumed in the model
are too high, for example, one will tend to overestimate the mass of
individual Machos and their density.  To study this, we measure the
event rate in the simulated galaxy and compare with the expected rate
assuming different models for the velocity distribution.

The visible parts of galaxies display substructure such as globular
clusters and dwarf galaxies.  Substructure is also likely to exist in
the distribution of dark matter especially in hierarchical clustering
models of galaxy formation.  Microlensing experiments are especially
sensitive to substructure in the Galaxy: If we are unlucky, a clump of
Machos will be passing between us and the LMC, biasing our estimates
for the density of Machos in the halo (Maoz 1994; Wasserman \&
Salpeter 1994; Metcalf \& Silk 1996, Zhao 1996).  Our
simulated microlensing experiment shows just such an effect.  Even
with our best-fit triaxial model, there are a few lines of sight in
which the measured optical depth is significantly higher than what is
predicted.  A close inspection of these lines of sight reveal that
substructure in the halo is often the cause of the discrepancy.

In Section II we survey the types of models previously considered for
the distribution of Machos.  The methods used to conduct microlensing
experiments in an N-body galaxy are developed in Section III.  Our the
results are presented in Section IV.  In Section V we describe how
hypothetical terrestrial dark matter detection experiments can be
performed in an N-body galaxy.  Section VI presents a summary of our
results and some concluding thoughts.

\section{Previous Analyses of the MACHO Experiment}

A model for Machos must specify their distribution in mass,
configuration space, and velocity space.  The ``standard'' halo model,
used as a benchmark by the MACHO collaboration, assumes that all of
the Machos have the same mass, that their velocities are isotropic and
Maxwellian, and that their density corresponds to that of a
cored isothermal sphere (Griest 1991; Alcock et al.\,1995, 1996).
This implies a distribution function (DF) of the form:

\be{dfmachos}
f({\bf x}, {\bf v}) = \rho({\bf x})F({\bf v})
\ee

\noindent where
\be{rho}
\rho({\bf x}) =
\frac{{\cal F} v_\infty^2}{4\pi G}
\frac{1}{r^2 + r_c^2}
\ee

\noindent and
\be{fofv}
F({\bf v}) = \frac{1}{\left (\pi v_\infty^2\right )^{3/2}}
\exp{\left (-\frac{v^2}{v_\infty^2}\right )}~.
\ee

\noindent $r_c$ is the core radius, $\cal F$ is the fraction of the halo
in Machos, and $v_\infty$ is the asymptotic circular speed of the
total halo.  In the standard model, $r_c=5\kpc$ and $v_\infty=220\kms$
while $\cal F$ and $M_0$ are left as free parameters.  It is a
likelihood analysis of this 2-parameter model that leads to the MACHO
collaboration's estimates for the mass and density of Machos in the
Galaxy.

The DF described above represents a highly idealized Macho halo.
Deviations between the model and the actual distribution of Machos may
introduce significant systematic errors in the analysis of a
microlensing experiment.  One can estimate these errors by considering
alternative models.  Several groups, for example, have considered
models in which the assumption of spherical symmetry is relaxed.  For
the most part these groups have focused on the seemingly reasonable
possibility that the halo is an axisymmetric oblate spheroid (Sackett
\& Gould 1993; Friemann \& Scoccimarro 1994; Gates, Gyuk, \& Turner
1995; Alcock et al.\,1995, 1996).  Models of this type can be
constructed by replacing $r^2$ in Eq.\EC{rho} by $m^2\equiv
x^2+y^2+z^2/q^2$ where $q$ is the axial ratio ($<1$ for oblate
spheroids) and by multiplying $\rho$ by $\lambda(q)\equiv
\sqrt{1-q^2}/\left ( q{\rm\, arccos{q}}\right )$.  Interestingly
enough, the optical depth and event rate do not change much as $q$
varies from $0.4$ (E6 oblate) to $1.0$ (spherical).  This is because
two competing effects cancel approximately.  As the model halo is made
more oblate, the central density must be increased if $v_\infty$ is to
be kept fixed.  This tends to increase the optical depth.  At the same
time our line of sight to the LMC passes through less halo material
which tends to reduce the optical depth.

The results for flattened halos have lead to the conjecture that the
total inferred mass of Machos within $50\kpc$ is relatively
independent of the assumed model ( Gates, Gyuk, \& Turner 1995;
Alcock et al. 1996).  However there is little direct evidence to
suggest that the dark Galactic halo is axisymmetric and oblate.
Observational clues about the shape of the dark Galactic halo come
from models of the metal-poor stellar halo (Gilmore, Wyse,
\& Kuijken 1989; van der Marel 1991), the outer satellites and
globular clusters (Hartwick 1996), and HI gas near the Galactic plane
(Merrifield \& Olling 1997).  The results are highly model dependent
and somewhat ambiguous leaving open the possibility that the Galactic
halo is prolate or triaxial.

Numerical experiments may provide the best hope for understanding the
structure of dark halos.  The halos found in N-body simulations of
dissipationless gravitational collapse are generally triaxial with a
slight preference to be prolate rather than oblate (Dubinski \&
Carlberg 1991; Warren et al.\,1992; see also the simulation discussed
below).  The inclusion of dissipational matter, which tends to settle
into a thin disk, will change these results somewhat typically driving
systems to become more oblate (Katz and Gunn 1991; Dubinski 1994)
though still generally triaxial.  And there are exceptions (Evrard,
Summers, and Davis 1994) suggesting that some halos may be
prolate-triaxial even when dissipational matter is included.  With
this in mind Holder \& Widrow (1996) have calculated the optical depth
and event rate for prolate halos.  For these models the central
density is lowered relative to what it would be in a spherical model.
This effect dominates so that predicted optical depth is reduced by a
significant amount.

Galactic microlensing experiments are also sensitive to the velocity
space distribution of the Machos.  If, for example, Machos are
preferentially on radial orbits, than the timescale for events would
be systematically longer than what would be expected assuming an
isotropic velocity distribution (Evans 1996).  The triaxial N-body
halos discussed above are supported by anisotropic velocity dispersion
which can affect the event rate and event duration (Holder \& Widrow
1996; Evans 1996).

One criticism of the models discussed above is that they do not
represent true equilibrium systems: with the exception of the special
case $r_c=0,~q=1$ (the singular isothermal sphere) a distribution
function given by Eqs.\,(1-3) does not satisfy the time-independent
collisionless Boltzmann equation.  Evans and Jijina (1994) have
attempted to address this concern by using the so-called power-law
models for the Macho DF.  These models are constructed from
simple power-law functions of the energy and angular momentum and
therefore automatically describe equilibrium systems.  Their main
advantage is that they are simple and analytic making lensing
calculations relatively easy.  They do have certain drawbacks.  First,
while the equipotential surfaces are spheroidal, the isodensity
surfaces are dimpled at the poles.  Second, once the disk is included,
the models no longer describe self-consistent equilibrium systems.

Finally, several groups have considered the implications of
subgalactic clustering in the distribution of Machos (Maoz 1994;
Wasserman \& Salpeter 1994; Metcalf \& Silk 1996; Zhao 1996).
In general these analyses make ad hoc assumptions about the
distribution and structure of Macho clusters.

\section{Preliminaries}

\subsection{The Simulated Halo}

Our halo is taken from a collisionless N-body simulation of a cold
dark matter (CDM) universe.  Of course Machos are, in all likelihood,
baryonic suggesting that a simulation of a baryon dominated universe
might be more appropriate.  However once Machos form they are
essentially collisionless.  We therefore expect that a Macho halo will
exhibit triaxiality, velocity space anisotropy, and substructure, much
like a CDM halo.

\begin{figure}
\plotone{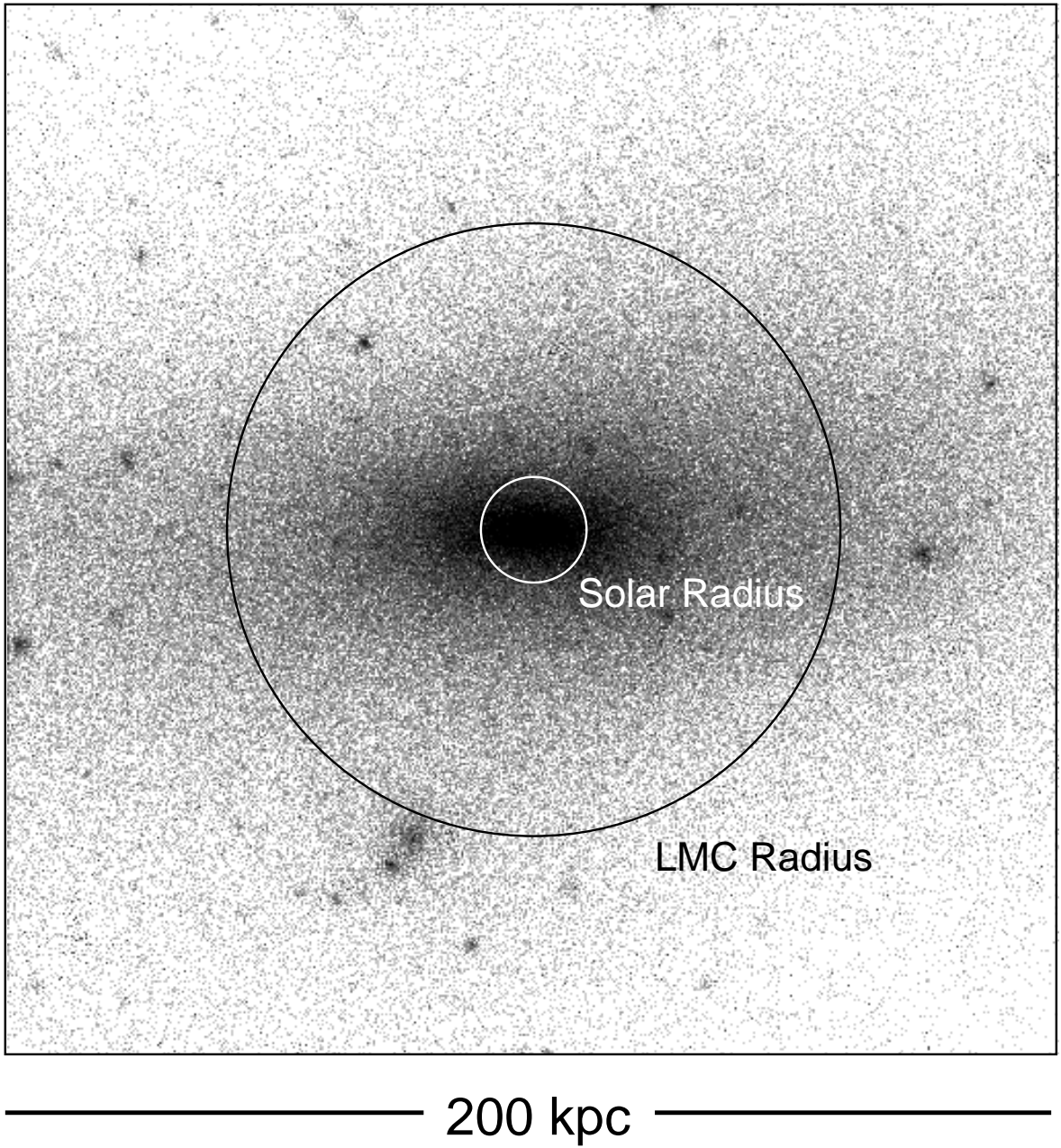}
\caption{ The N-body galaxy in projection.  This
projection is along the intermediate axis of the galaxy so that its prolate
nature is readily apparent.  The inner circle represents the
positions of hypothetical observers; the outer circle
represents the positions of hypothetical LMCs.}
\label{fig-nbody}
\end{figure}

The halo is extracted from cosmological collapse simulation run with a
parallel treecode on the Pittsburgh Cray T3E (Dubinski 1996).  The
halo is comprised of roughly 700,000 particles within the virial
radius though there are only 260,000 particles within the radius of
the LMC (Figure~\ref{fig-nbody}).  Each particle has a mass $M_{\rm P}
= 2.7\times 10^{6}\,M_\odot$.  The experiments are conducted at two
different epochs corresponding to ages of $9.7\,{\rm Gyr}$ and
$16.0\,{\rm Gyr}$.  The results are essentially the same for the two 
time frames and we
present only those from the latter.  The galaxy is prolate-triaxial
and has a relatively flat rotation curve out to large radii (Figure
2).  We model the mass distribution of the galaxy by assuming that
isodensity surfaces are ellipsoidal so that $\rho=\rho(m)$ where $m^2
= x^2 + y^2/q_1^2 + z^2/q_2^2$ provided we have chosen the axes to
coincide with the principle axes of the Galaxy.  The density profile
is fit to an NFW profile (Navarro, Frenk, \& White 1996):

\be{nfw}
\rho(m)~=~\frac{M_s}{4\pi q_1 q_2}\,\frac{1}{m\left (m + a_s\right )^2}
\ee

\noindent where $M_s$ is equal to the mass inside an ellipsoidal radius
$m=3a_s$.  $M_s,\,a_s,\,q_1,$ and $q_2$ determined directly from the
simulation.  We also use less sophisticated models, specifically, a
spheroidal model ($q_1=q_2=0.5$) and a spherical one.  The model
parameters are summarized in Table 1.

\begin{table}
\begin{center}
\begin{tabular}{|| c | c | c | c | c ||}\hline
~~~~Model~~~~& $q_1$ & $q_2$ & $M_s (10^{11}M_\odot) $ & $a_s (\kpc)$ \\ \hline
I (triaxial)    & 0.55 & 0.44 & 5.2 & 21.4 \\ \hline
II (spheroidal)   & 0.5  & 0.5  & 5.2 & 21.2 \\ \hline
III (spherical)  & 1.0  & 1.0  & 6.5 & 15.9 \\ \hline
\end{tabular} 
\end{center}
\caption{Model parameters}
\end{table}

\begin{figure}
\plotone{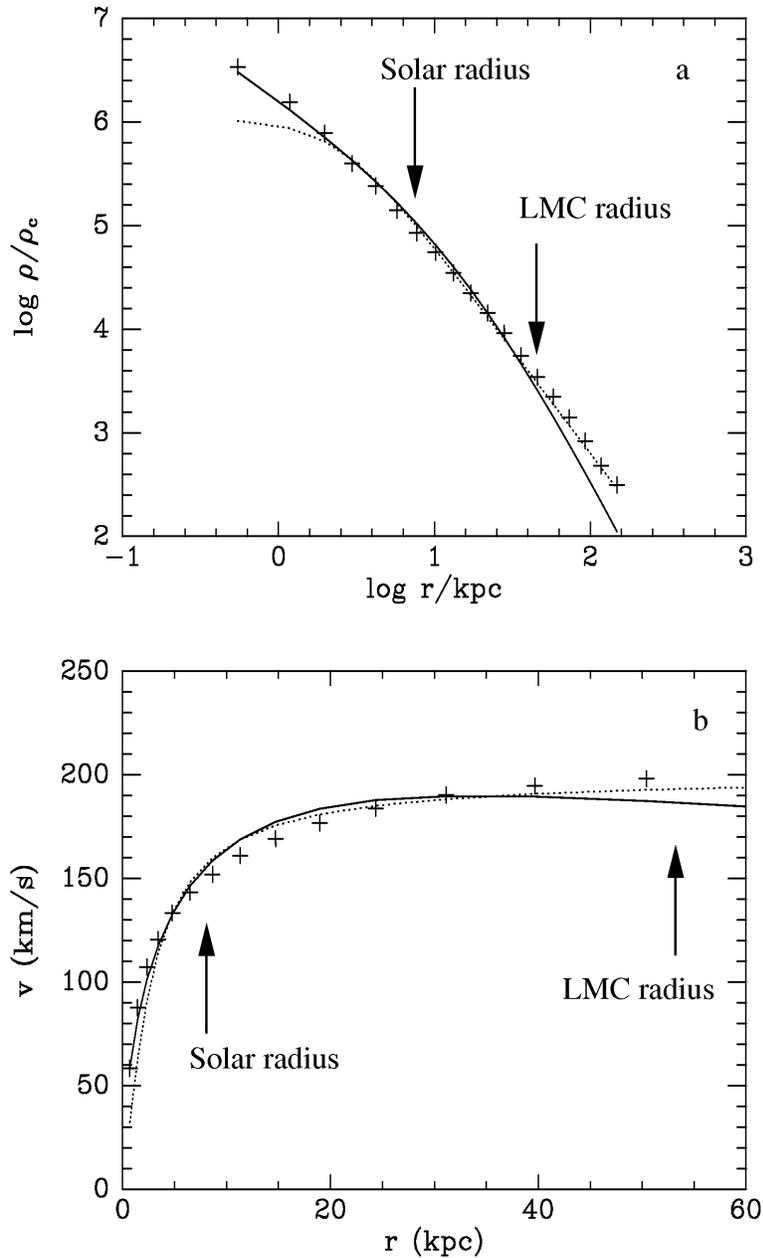}
\caption{(a) Density profile and (b) rotation curve for the 
simulated halo.  Crosses represent the spherically averaged density
and circular rotation speed for the halo.  The solid lines are for
the NFW profile discussed in the text.  The dotted lines are for the
best-fit cored isothermal sphere ($r_c=2.4\kpc$, $v_\infty=200\kms$).}
\label{fig-rot}
\end{figure}

$M_s$ and $a_s$ depend sensitively on the region of the halo used in
the fit.  In other words, Eq.\,\EC{nfw} does not provide a very good
global model for this particular galaxy.  This is especially true for
the $16\,{\rm Gyr}$ time frame where the halo is merging with a
satellite.  Of course, microlensing experiments only probe the region
of the halo between the observer and the target stars (i.e., between
$8.5\kpc$ and $50\kpc$) where the density profile is close to a power
law.  The parameters in Table 1 are chosen to fit the density profile
in this region.  In fact, any number of fitting formulae could be
used.  For example, a cored isothermal sphere (Eq.\,\EC{rho}) with
$r_c=2.4\kpc$, ${\cal F}=1$, and $v_\infty=200\kms$ does just as well since
the observer is outside the core radius and the LMC is inside the
radius where the density profile becomes steeper than $r^{-2}$.  In
Figure~\ref{fig-rot} we show the circular rotation speed 
$v_c\,\left (\equiv
(GM(r)/r)^{1/2}\right )$ for the halo, Model I, and the ``best-fit''
cored isothermal sphere.

A galactic coordinate system is set up in the usual way.  Our
hypothetical observer is at the origin, the center of the galaxy is on
the $y$-axis a distance $r_s =8.5\kpc$ away, and the $z$-axis points
``north''.  The hypothetical LMC is at galactic coordinates $(l,\,b) =
(-32.8^\circ,\,281^\circ)$ a distance $L$ from the origin.  For
simplicity we assume that the hypothetical LMC, in projection as seen
from the observer, is a circle of radius $R_{\rm LMC}$.  The
foreground of the LMC is a conical volume $\Omega$ with base diameter
$2R_{\rm LMC}$ and height $L$.

Different realizations of the experiment are performed by varying the
orientation of the halo keeping the observer, LMC, and center of the
halo fixed.  Since our halo does not have a disk, the observer is not
constrained to any particular plane.  We therefore have three degrees
of freedom (the Euler angles) in choosing the orientation of the halo.
Equivalently, we can imagine keeping the halo fixed and considering a
family of hypothetical observers positioned on the surface of a sphere
a distance $r_s$ from the center of the halo.  The hypothetical LMCs
lie on a concentric sphere of radius $\left (L^2 -2r_s
L\cos{b}\cos{l}+r_s^2 \right )^{1/2}$ (Figure~\ref{fig-nbody}).  For
each observer, there is a circle of LMCs a distance $L$ away that have
the correct orientation.

\subsection{Optical Depth}

The optical depth $\tau$ is the probability that light from a star in
the LMC will be amplified by a factor $A$.  This requires that a Macho
pass within a distance $R=uR_{\rm E}$ from the line of sight where $u =
2^{1/2}\left [A\left (A^2 -1\right )^{-1/2} -1\right ]^{1/2}$, $R_{\rm E}$
is the Einstein radius:

\be{ER}
R_{\rm E}(z') = 9.65\times 10^{-8}\left (
\frac{M}{M_\odot}\frac{L}{50\kpc}
\frac{z'\left (L - z'\right )}{L^2}\right )^{1/2}\kpc~,
\ee

\noindent and $z'$ is the distance to the Macho.  For definiteness we
set $A=1.34~ (u=1)$ and $L=50\kpc$.  If all of the Machos have the
same mass then $\tau$ is equivalent to the number of Machos in a tube
(the so-called lensing tube) whose axis is the line of sight and whose
radius is $R_{\rm E}$.  The expectation value of $\tau$, given an analytic
model for $f$, is

\bea{tau}
\langle\tau\rangle
     & = & \frac{1}{M}\int d^3{\bf x}\rho({\bf x}) \\
     & = & \frac{1}{M}\int_0^L dz'\rho(z')\pi R_{\rm E}^2(z')
\eea

\noindent where $\rho({\bf x})=\int d^3 {\bf v} f({\bf x},{\bf v})$.
We also calculate $\sigma_\tau$, the standard deviation one expects
for $\tau$ assuming particles are chosen at random from the model DF.
The relevant equations for this are derived in the Appendix.

We can measure $\tau$ in our N-body halo by counting the number of
particles in a lensing tube whose radius is given by Eq.\,\EC{ER} with
$M=M_{\rm P}$.  (Recall that, for fixed mass density in Machos, $\tau$
is independent of $M$.)  Of course only a very small fraction of the
tubes will contain a particle.  We can mimic what is done in the MACHO
experiment by counting the number of particles in a large number
$(\sim 10^6)$ of tubes each of which lies, more or less, in the volume
$\Omega$.  However, it is more efficient to use all of the particles
in $\Omega$, weighting them by an appropriate geometric factor.  To be
precise we write the DF for our N-body galaxy as the sum of
$\delta$-functions:

\be{nbodydf}
f = M_{\rm P}\sum_i \delta\left ({\bf x}-{\bf x}_i\right )
 \delta\left ({\bf v}-{\bf v}_i\right )
\ee

\noindent where $i$ labels the particles in the simulation.
Substituting into Eq.\,\EC{tau} we have

\be{nbodytau}
\tau = \sum_i\left (\frac{R_{\rm E}(z_i)}
{\theta_{\rm LMC}z'_i}\right )^2
\ee

\noindent where the sum is over all particles in $\Omega$
and we use $M=M_{\rm P}$ in calculating $R_{\rm E}$.

An alternative strategy is to replace $\Omega$ with a volume
$\tilde\Omega$ whose shape is the same as that of a lensing tube.  For
example, we can choose $\tilde\Omega$ to be a tube of radius
$R_{\rm tube}(z') =2R_{\rm LMC}\left (z'(L-z')/L^2\right )^{1/2}$ so
that the maximum radius of the tube is equal to the radius of the LMC.
In this case all particles contribute equally to $\tau$ making it easy
to visualize the phase space distribution of objects in the tube.
Furthermore, near the observer, where the density of halo objects is
presumably highest, $\tilde\Omega$ is fatter than $\Omega$. We
therefore expect more objects in the tube and hence better statistics.
The main disadvantage of this strategy is that it distorts the
geometry of a realistic microlensing experiment.  The angular diameter
of $\tilde\Omega$ is $\tilde\theta=2\tan^{-1}\left ((L-z')/z'\right )$
which is far too big near the observer.  We will therefore use
$\Omega$ to calculate the optical depth.  However we use
$\tilde\Omega$ to calculate the event rate where it is difficult to
estimate statistical uncertainties when $\Omega$ is used (see Appendix
and the next section).

\subsection{Event Rate}

In practice the expected number of events, $N_{\rm exp}$, is more
useful than the optical depth when comparing predictions with
observations.  $N_{\rm exp}$ is given by

\be{nexp}
N_{\rm exp} = E\int_0^\infty
\frac{d\Gamma}{d\th}\epsilon(\th) d\th
\ee

\noindent where $\th = 2R_{\rm E}/v_\vp$ is the event duration, $v_\vp$
is the magnitude of the transverse Macho velocity in the lensing tube
rest frame, $\epsilon(\th)$ is the detection efficiency, and $E$ is
the ``exposure'' for the experiment given in units of ``star-yr''.
The differential event rate is given by

\be{gamma}
d\Gamma = \frac{1}{M} d\sigma d{\bf v}f({\bf x},{\bf v}) 
v_\vp \cos{\theta} 
\ee

\noindent where $d\sigma = R_{\rm E} d\phi dz'$ is a surface element on 
the lensing tube and $d^3 {\bf v} = dv_{z'}v_\vp dv_\vp d\theta$.
Given an analytic function for $f$, the total event rate $\Gamma$ is
found by integrating over ${\bf v}$ and $\sigma$.  To calculate
$d\Gamma/d\th$, the event rate as a function of event duration, we change
variables from $v_\vp$ to $\th$, differentiate with respect to $\th$,
and integrate over the remaining variables.  

If the velocities of the Machos are isotropic and 
Maxwellian (Eq.\,\EC{fofv}) than the expectation values for $\Gamma$ 
and $d\Gamma/d\th$ are

\be{gammaiso}
\langle\Gamma\rangle = 
\frac{\pi^{1/2}v_\infty}{M}\int_0^L dz'\rho(z')R_{\rm E}(z')
\ee

\noindent and

\be{dgdt}
\left\langle\frac{d\Gamma}{d\th}\right\rangle =
\frac{32}{\th^4 M v_\infty^2}
\int_0^L \rho(z')R_{\rm E}(z')\exp{\left [
-\frac{rR_{\rm E}^2(z')}{\th^2 v_\infty^2}\right ]}dz'~.
\ee

\noindent An expression for $\sigma_\Gamma$, the standard deviation 
for $\Gamma$ assuming particles are chosen at random from an analytic
DF, is derived in the appendix.

Eq.\,\EC{gamma} represents a flux of particles through a surface,
something that is difficult to measure in an N-body simulation.  To
circumvent this problem we multiply both sides of Eq.\,\EC{gamma} by
$\varpi d\varpi$ where $\vp$ is the distance from the axis of the
lensing tube:

\be{gamma_volume}
\vp d\vp d\Gamma = \frac{1}{M_{\rm P}} 
d^3{\bf x} d^3{\bf v}f({\bf x},{\bf v}) 
v_\vp \cos{\theta} R_{\rm E}(z')~.  
\ee

\noindent Substituting Eq.\,\EC{nbodydf} for $f$ and 
integrating over the volume $\tilde\Omega$ yields an expression for
$\Gamma$ that is directly applicable to an N-body galaxy:

\be{nbodygamma}
\Gamma = \frac{2}{\pi}\sum_i\frac{R_{\rm E}(z_i) v_{\vp,i}}
{R_{\rm tube}^2(z_i)}~.
\ee

\noindent $d\Gamma/d\th$ is estimated by binning the 
events according to event duration.

\subsection{An Illustrative Example}

We illustrate the techniques developed above in a toy galaxy where the
positions and velocities of the particles are chosen at random from a
known analytic DF.  For definiteness we assume that this DF is given
by Eqs.(1-3) with ${\cal F}=1$, $v_\infty=200\,\kms$, and $r_c=5\kpc$.
The galaxy is represented by 200,000 $M_{\rm P}=2.2\times
10^{6}M_\odot$ mass particles.  1000 individual microlensing
experiments are performed, each of which assumes a different
orientation for the galaxy.  To be precise, let
$(\alpha,\,\beta,\,\gamma)$ be the three Euler angles which describe
the galaxy's orientation.  We take $10$ even steps in $\alpha$ (from
$0$ to $2\pi$), $\cos{\beta}$ (from $-1$ to $1$), and $\gamma$ (from
$0$ to $2\pi$).  We choose $R_{\rm LMC}=4\kpc$.  This is roughly a
factor of two larger than the real LMC, a choice made to increase the
number of particles in the experimental volume $\Omega$.

From the analytic DF we calculate $\langle\tau\rangle$,
$\langle\Gamma\rangle$, and $\langle d\Gamma/d\th\rangle$ as well as
$\sigma_\tau$ and $\sigma_\Gamma$.  We next compute, for each
experiment, the normalized errors

\be{nerrortau}
\varepsilon_\tau~=~\frac{\tau - \langle\tau\rangle}{\sigma_\tau}
\ee

\be{nerrorGamma}
\varepsilon_\Gamma~=~\frac{\Gamma - \langle\Gamma\rangle}{\sigma_\Gamma}~.
\ee

\begin{figure}
\plotone{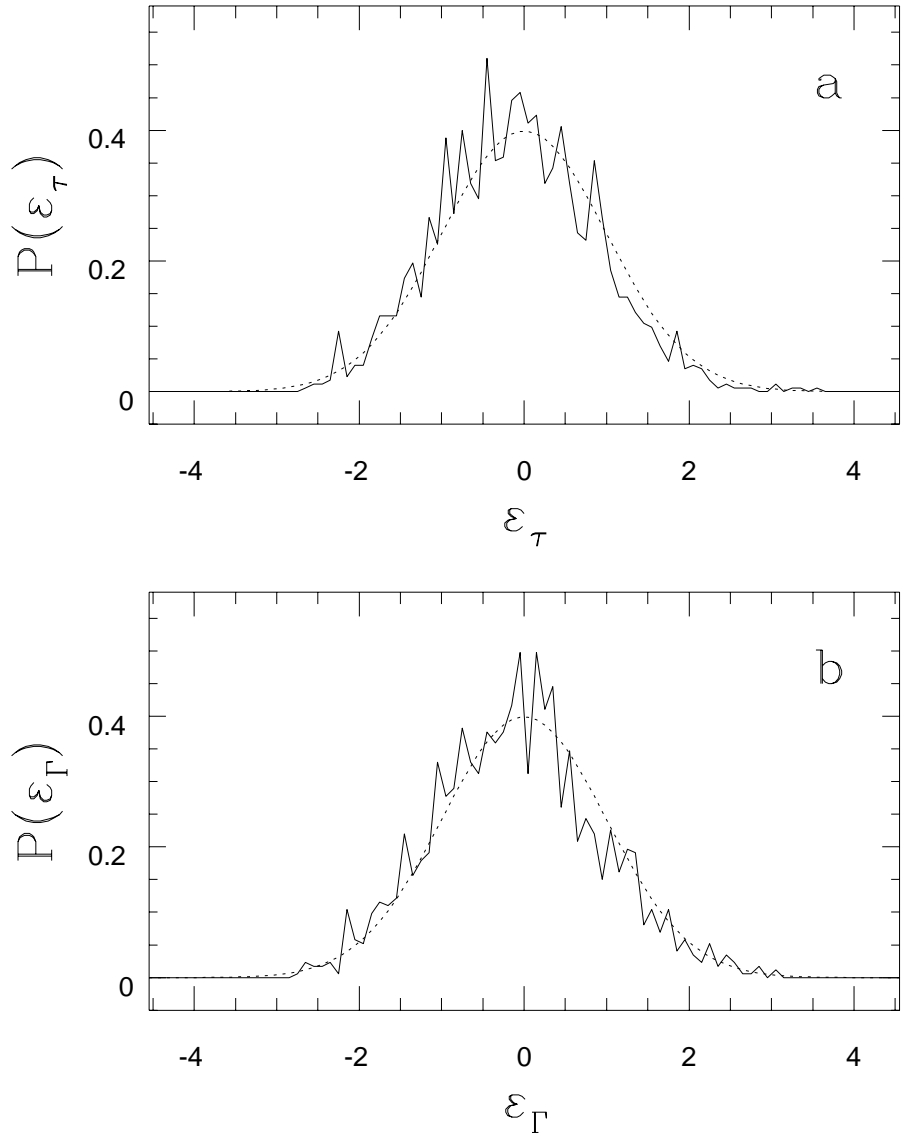}
\caption{ Results of hypothetical microlensing experiments
in the ``toy'' model galaxy.  (a) $P(\varepsilon_\tau)$ as a function
of $\varepsilon_\tau$; (b) $P(\varepsilon_\Gamma)$ as a function of
$\varepsilon_\Gamma$.  The solid lines are the measured distributions 
for the 1000 experiments.   The dotted lines are Gaussians of unit
variance. }
\label{fig-pepsgam}
\end{figure}

\noindent Figures~\ref{fig-pepsgam}a and \ref{fig-pepsgam}b 
give the probability distributions 
$P(\varepsilon_\tau)$ and $P(\varepsilon_\Gamma)$ for the 1000
experiments.  $P(\varepsilon_\tau)$ and $P(\varepsilon_\Gamma)$ are each
well-approximated by a Gaussian of unit variance as they should given
that the particles are drawn at random from the same distribution used
in the model calculations.

It is also useful to calculate the fractional error

\be{ferrortau}
\mu_\tau~=~\frac{\tau - \langle\tau\rangle}{\langle\tau\rangle}
\ee

\be{ferrorGamma}
\mu_\Gamma~=~\frac{\Gamma - \langle\Gamma\rangle}
{\langle\Gamma\rangle}~.
\ee

\noindent $\mu_\tau$ and $\mu_\Gamma$ translate directly to the fractional
error in the hypothetical observer's estimate for the mass fraction in
Machos.  In this example, where the model is spherically symmetric,
$\langle\tau\rangle$, $\langle\Gamma\rangle$, $\sigma_\tau$, and
$\sigma_\Gamma$ are the same in each experiment.  We can therefore
obtain $P(\mu)$ from $P(\epsilon)$ simply by substituting
$\mu_\tau=\epsilon_\tau\sigma_\tau/\langle\tau\rangle$ and
$\mu_\Gamma=\epsilon_\Gamma\sigma_\Gamma/\langle\Gamma\rangle$.  The
situation is more complicated when an aspherical model is used since
$\langle\tau\rangle$, $\langle\Gamma\rangle$, $\sigma_\tau$, and
$\sigma_\Gamma$ will then depend on the orientation of the galaxy.  In
the analysis below, we use $\epsilon_\tau$ and $\epsilon_\Gamma$ to
illustrate deviations from Gaussian statistics and $\mu_\tau$ and
$\mu_\Gamma$ to indicate the potential errors one might encounter in
determining the mass fraction of Machos.

\begin{figure}
\plotone{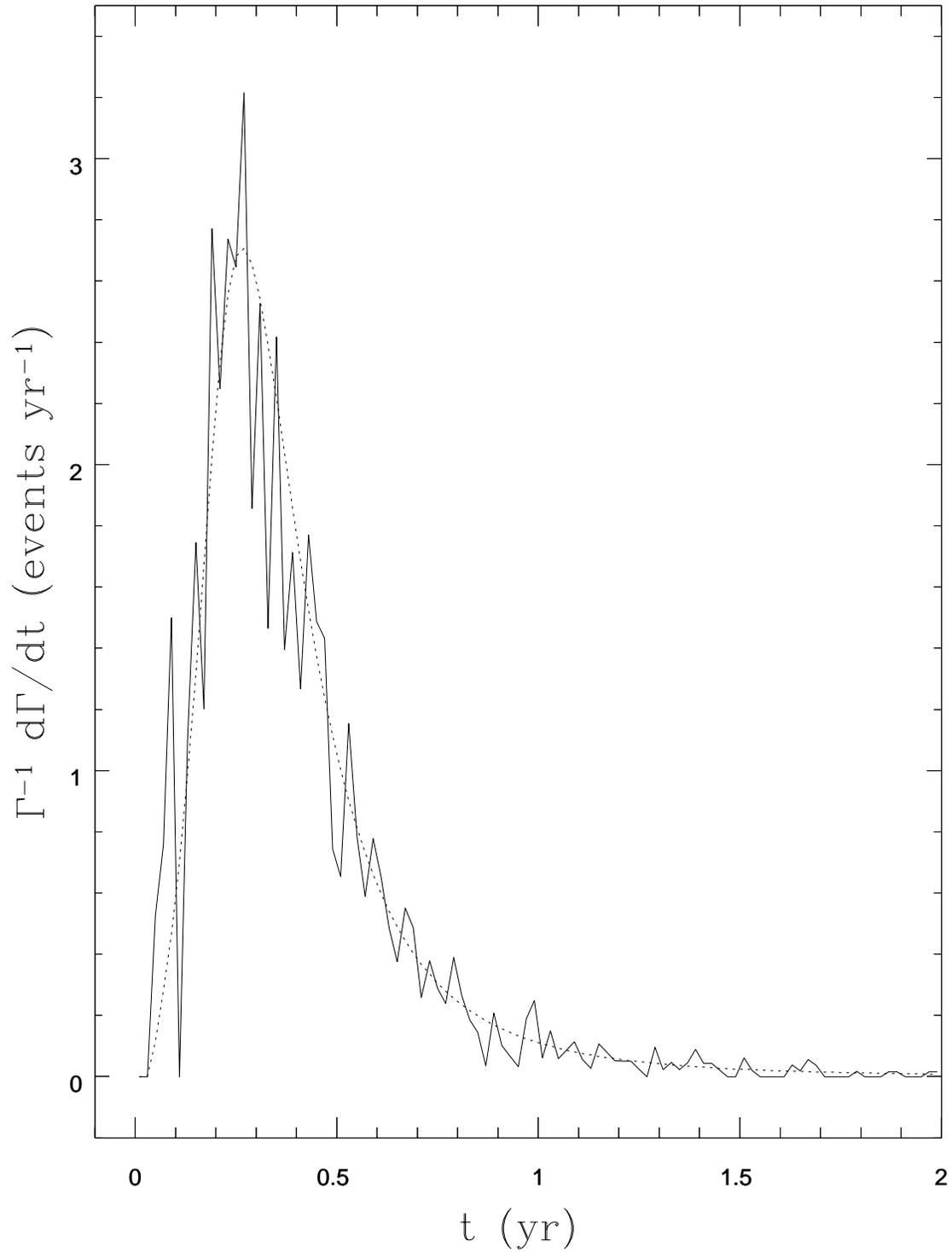}
\caption{ Event rate as a function of event duration for a 
particular microlensing experiment.  The solid line gives the measured
$\Gamma^{-1}d\Gamma/d\th$ as a function of $\th$.  The dotted line is
the model prediction.}
\label{fig-dgamdt}
\end{figure}

Figure~\ref{fig-dgamdt} shows the measured $d\Gamma/d\th$ 
for one of the experiments as compared with the predicted $d\Gamma/d\th$ 
from Eq.\,\EC{dgdt}.

\section{Results}

\begin{figure}
\plotone{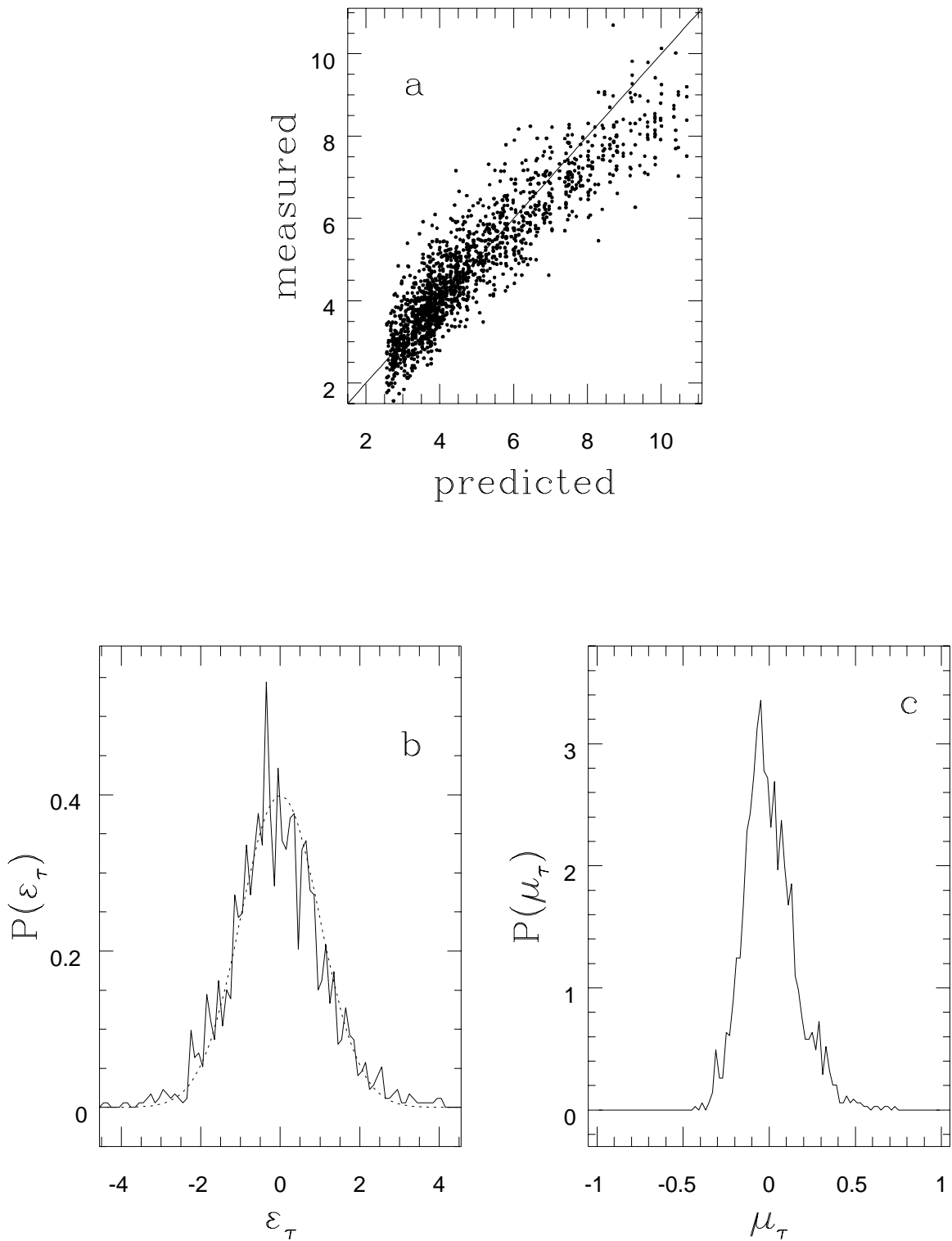}
\caption{ (a) Measured versus predicted optical depth
$\tau$ using the ``best-fit'' triaxial model for the galaxy.  Each
point represents a different microlensing experiment
(i.e., different orientation of the simulated galaxy).  $\tau$ is
given in units of $10^{-7}$.  There are 1728 experiments in total.  In
a typical experiment, $100-200$ particles 
contribute to the optical depth. (b) $P(\varepsilon_\tau)$ as a
function of $\varepsilon_\tau$. (c) $P(\mu_\tau)$ as a function of
$\mu_\tau$.}
\label{fig-triax}
\end{figure}

$12^3=1728$ microlensing experiments are performed in the N-body halo
described in Section 3.1.  The optical depth and event rate are
measured and compared with predictions made assuming one of the three
models of Table 1.  The results for our best-fit
triaxial model (Model I) are presented in
Figures~\ref{fig-triax}a, \ref{fig-triax}b, and
\ref{fig-triax}c.  Figure~\ref{fig-triax}a is a scatter
plot of measured versus predicted $\tau$ for the $1728$ experiments.
The probability distribution $P(\varepsilon_\tau)$ for the normalized
error is reasonably well approximated by a Gaussian of unit variance
(Figure~\ref{fig-triax}b).  There are, however, two experiments where
the deviations between measured and predicted $\tau$ differ by more
than $4\sigma_\tau$ and $14$ experiments where the deviations differ
by more than $3\sigma_\tau$, significantly more than what is expected
from Gaussian statistics.  The fractional error $\mu_\tau$ in most of
the experiments is $\la 0.3$ (Figure~\ref{fig-triax}c) and comes
primarily from statistical fluctuations (cf. Figure~\ref{fig-triax}b).
However, in the rare experiments where $\varepsilon_\tau\ga
3\sigma_\tau$, the fractional error can be as high as $0.7$.

\begin{figure}
\plotone{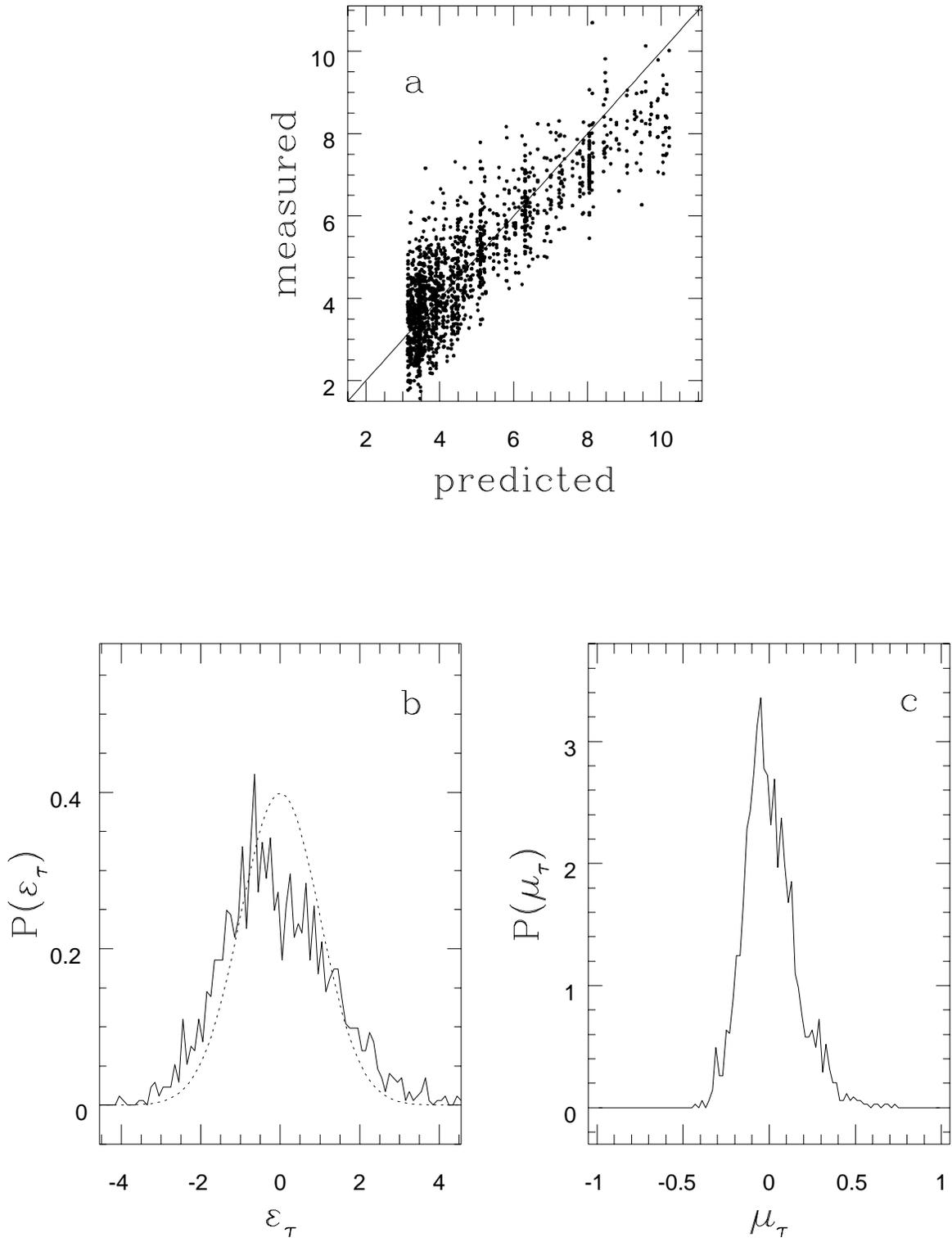}
\caption{ Same as Figure~\ref{fig-triax} but for Model II.}
\label{fig-spheroid}
\end{figure}

Figures~\ref{fig-spheroid} presents the results for the spheroidal
model (II).  Because of the symmetry in the model, different
orientations of the galaxy can lead to identical predictions.  This
accounts for the vertical stripes in Figure~\ref{fig-spheroid}a.
Model II is fairly close to the triaxial model: The axial ratios $q_1$
and $q_2$ have changed by roughly $10\%$.  Nevertheless, there are
noticeable systematic effects, as can be seen by comparing
Figures~\ref{fig-spheroid}b and
\ref{fig-spheroid}c with \ref{fig-triax}b and \ref{fig-triax}c.

\begin{figure}
\plotone{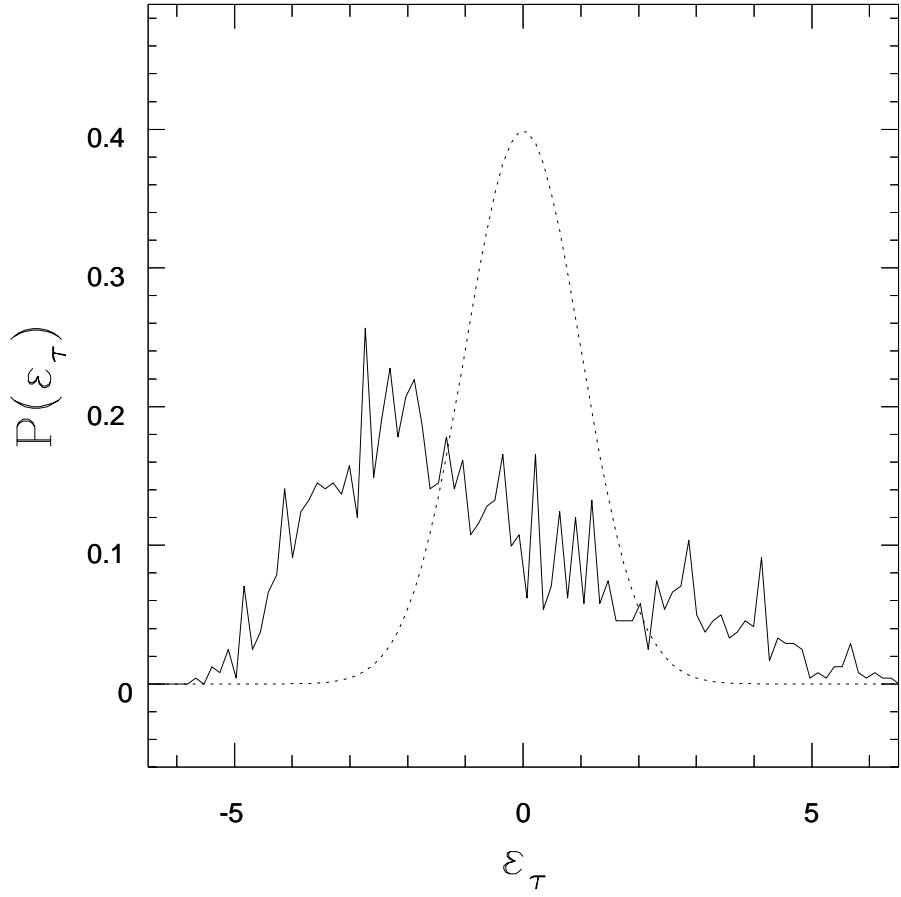}
\caption{ $P(\varepsilon_\tau)$ versus $\varepsilon_\tau$
for Model III.  In this case, the predicted $\tau$ is the same for all
experiments.  It is therefore easy to translate this figure into a
probability distribution for either $\tau$ or $\mu_\tau$ (see
text).}
\label{fig-sphere}
\end{figure}

The results for the spherical model (III) are presented in
Figure~\ref{fig-sphere}.  As in the toy model discussed above, the
predicted quantities are the same for each of the experiments.  We
show the normalized error distribution $P(\epsilon_\tau)$.  We can
obtain the probability distributions for $\tau$ and $\mu_\tau$ through
the relations $\tau = \sigma_\tau\varepsilon_\tau +
\langle\tau\rangle$ and $\mu_\tau =
\sigma_\tau\varepsilon_\tau/\langle\tau\rangle$ where $\sigma_\tau =
6.54\times 10^{-8}$ and $\langle\tau\rangle= 5.31\times 10^{-7}$.
In a large number of experiments the model overestimates the optical
depth ($\varepsilon_\tau<0$).  These correspond to orientations
of the halo where the long axis is near the ``galactic plane''.
Likewise, experiments where the model underestimates $\tau$ find the LMC
close to the long axis.  (This is the orientation considered by
Holder \& Widrow (1996).)  On the whole, agreement between the measured
and predicted optical depths is rather poor:  The rms normalized error
$\sqrt{\bar{\varepsilon_\tau^2}}=2.65$ corresponding to an rms
fractional error of $\sqrt{\bar{\mu_\tau^2}}=0.33$.

Figure~\ref{fig-gamma-triax} presents results for the total event rate
$\Gamma$.  The triaxial model is used for the mass distribution and
the velocities are assumed to be isotropic and Maxwellian
(Eq.\,\EC{fofv}).  For this model we take $v_\infty=212\kms$ which is
found by computing the rms velocity for all of the particles in the
halo.

The experiments with high event rate correspond to orientations of the
galaxy which put the LMC on the long axis and the observer in the
plane containing the two short axes.  In these experiments, the model
overestimates the event rate by as much as $20\%$.  As discussed in
Section 3.1, the parameters in the NFW profile depend sensitively on
the region of the galaxy used in the fit.  The fits in Table 1 are
based on the region of the halo inside an ellipsoidal radius of
$50\,{\rm kpc}$.  However, the high $\Gamma$ experiments probe smaller
ellipsoidal radii where the model tends to overestimate the density.
A similar problem might arise in actual microlensing experiments since
the density profile for the Galaxy is determined from observations of
stars and gas near the Galactic disk while the LMC lies close to the
south Galactic pole.

The triaxial structure of our simulated galaxy is supported by
anisotropic velocity dispersion.  Indeed the velocity dispersion in
the $x$-direction is a factor $1.25$ higher than in the $y$ and $z$
directions, in agreement with what one expects from the tensor virial
theorem for a galaxy of this shape.  A simple prescription for taking
velocity anisotropy into account has the velocity distribution in
Eq.\,\EC{fofv} replaced by a modified Maxwellian function of the form
(Holder \& Widrow 1996):

\be{modmax}
F({\bf v}) = \frac{1}{\left (\pi \xi_1 \xi_2 v_\infty^2\right )^{3/2}}
\exp{\left (-\frac{v_x^2}{v_\infty^2}
-\frac{v_y^2}{\xi_1^2v_\infty^2} - 
\frac{v_z^2}{\xi_2^2v_\infty^2}\right )}~.
\ee

\noindent $\xi_1$ and $\xi_2$ are related to $q_1$ and $q_2$
through the tensor virial theorem.  While this model still makes the
rather dubious assumption that the distribution of Machos in velocity
space is independent of position it does connect the shape of the halo
with its velocity space structure in a reasonable way.
Unfortunately calculating model predictions with Eq.\,\EC{modmax} is
quite cumbersome.  We choose instead to consider a model in which the
velocity distribution is assumed to be isotropic but with $v_\infty$
determined separately for each experiment using the particles in the
``experimental'' volume $\tilde\Omega$.  The results, presented in
Figure~\ref{fig-gamm-improve}, show a clear improvement over those in
Figure~\ref{fig-gamma-triax} where the same $v_\infty$ was used for
all experiments.  While ad hoc, this comparison suggests that
velocity anisotropy can lead to systematic errors in a microlensing
experiment.  To further illustrate this point we measure
$d\Gamma/d\th$ for a single experiment and compare with model
predictions.  This is done in Figure~\ref{fig-dgamdt-anios}.  Two
models are chosen, one with $v_\infty$ taken from the simulation as a
whole, and the other with $v_\infty$ measured locally (i.e., in the
volume $\tilde\Omega$).  Clearly the latter provides better agreement
between theory and experiment though the net effect on $N_{\rm exp}$ 
(which is calculated by integrating over $\hat t$) may
be rather small.

As noted above, even with the triaxial model, the number of
experiments with a relatively high ($\varepsilon_\tau >3\sigma$)
discrepancy between measured and observed optical depth is greater
than expected had the particles been chosen at random from the model
DF.  We have looked in detail at the phase space distribution of
particles in these lensing tubes.  (Here, we use the volume
$\tilde\Omega$ which has the same shape as the lensing tube.  This
avoids the complication of having particles of different weight in the
sum for $\tau$.)  In Figure~\ref{fig-vp-vz}a we show the distribution
of particles as a function of $v_\vp$ and $z'$, the two relevant phase
space coordinates for microlensing experiments.  A clump of
approximately $50$ particles ($M_{\rm clump}\sim 10^8 M_\odot$), 2/3
of the way to the LMC and moving through the lensing tube with a
velocity of $450\,{\rm km/sec}$, is clearly visible.  With enough
events, one might ``see'' such a clump in a plot of the event rate as
a function of event duration (Figure~\ref{fig-vp-vz}b).  However, the
clump is not nearly so pronounced in an ``observer's-eye'' view of
this region of the sky (Figure~\ref{fig-vp-vz}c).

Substructure of this type is entirely expected in hierarchical
clustering models such as CDM.  The object shown in
Figure~\ref{fig-vp-vz} is a dwarf halo that will spiral, under the
influence of dynamical friction, toward the center of the main halo,
eventually being stripped apart by tidal interactions.  Objects of this
type arise in about $1\%$ of our experiments, a small but still
significant fraction.  Of course substructure in the stellar halo of
the Galaxy has already been observed (see, for example, Majewski,
Munn, \& Hawley 1994; Ibata, Gilmore, \& Irwin 1995).  Indeed,
observations by Zaritsky \$ Lin (1997) indicate that there may be an
excess of stars between us and the LMC.  Though these observations are
somewhat controversial (Alcock et al.\,1997a) they do suggest an
alternative explaination for MACHO's results: microlensing by
foreground stars of a ``lumpy'' halo (Zaritsky \& Lin 1997; Zhao
1997).

\section{WIMP and Axion Search Experiments in a Simulated Galaxy}

There are currently a large number of experiments, either in operation
or in construction, designed to search for elementary particle dark
matter candidates such as weakly-interacting massive particles (WIMPs)
and axions (Dougherty 1995; Rosenberg 1995).  Typically these
experiments measure the density and kinetic energy distribution of
dark matter particles passing through a terrestrial laboratory.  The
results of such experiments are therefore subject to systematic
effects similar to the ones discussed in the previous sections.
Clearly, the large scale structure of the halo, such as its shape,
density profile, and velocity space structure, will determine the local
density and energy distribution of dark matter particles.  Small scale
structures such as those found in our microlensing experiments can
also affect the outcome of dark matter search experiments.  Along
similar lines, Sikivie, Tkachev, and Wang (1996) have suggested that
there may be discrete peaks in the local energy distribution of dark
matter particles.  Their analysis is based on the secondary infall
model of Fillmore \& Goldreich (1984) and Bertschinger (1985) which is
characterized by an intricate fine-grained structure in phase space.
This model assumes spherical symmetry, radial orbits, and smooth
accretion of matter and no doubt provides a highly idealized
description of the Galactic halo.  Still the work raises the
interesting possibility that phase space structure can have an impact on
dark matter search experiments.

We can test some of these possibilities by performing dark matter
search experiments in an N-body halo.  We measure the kinetic energy
distribution of particles in a volume $V$ centered on a hypothetical
observer and compare the results with model predictions.  Let ${\bf
x}_s$ be the position vector of a hypothetical observer as measured
from the center of the halo, $\rho({\bf x}_s)$ be the density of dark
matter in the region of the observer, and $F({\bf v})$ be the local
velocity distribution function, i.e., $f({\bf x}_s, {\bf v})\equiv
\rho({\bf x}_s)F({\bf v})$.  The number of particles with kinetic
energy per unit mass between $\kappa$ and $\kappa+\Delta\kappa$ is

\bea{dndk}
N(\kappa)
     & = & \frac{dn}{d\kappa}\Delta\kappa \\
     & = & \frac{\rho({\bf x}_s)V\sqrt{2\kappa}}\Delta\kappa{M_P}
\int F({\bf v})d\cos{\theta_v} d\phi_v
\eea
 
\noindent where $d^3{\bf v}=v^2 dv \, d\cos{\theta_v}\,d\phi_v=
\sqrt{2\kappa} d\kappa \,d\cos{\theta_v}\,d\phi_v$.
As an illustrative example, we perform $100$ dark matter search
experiments taking $V$ to be a sphere of radius $2\kpc$.
Figure~\ref{fig-wimp}a shows the results for the total number of
particles $N_V$ in the volume $V$.  Each point in the plot displays
the measurement and prediction for a different observer.  The model
used in making the predictions is the same triaxial model discussed
above.  In addition, we assume that the velocities are isotropic and
Maxwellian.  Figure~\ref{fig-wimp}b shows normalized energy spectra as
measured by 4 different observers.  The results are consistent with
what one would expect from statistical fluctuations.

The difficulty we face is that with only $700,000$ particles in the
halo, the spatial and energy resolution is rather poor.  A detailed
energy spectrum would require $N_V\gg 100$ while the structures we are
interested in may well have a scale significantly less than a
kiloparsec.  Unfortunately, for this simulation, the number of particles in
a sphere of radius $R$ is $N_V\sim 100\,(R/1\kpc)^3$.

\section{Conclusions and Caveats}

The dark Galactic halo is, in all likelihood, a very complicated
system.  Most theories of galaxy formation would predict that it
is triaxial in shape and that the velocity distribution varies
from point to point and is generally anisotropic.  Moreover, if the
Galactic halo is built up from smaller collapsed objects, as in
hierarchical clustering models, it will contain substructure on a
wide range of subgalactic scales.

\begin{figure}
\plotone{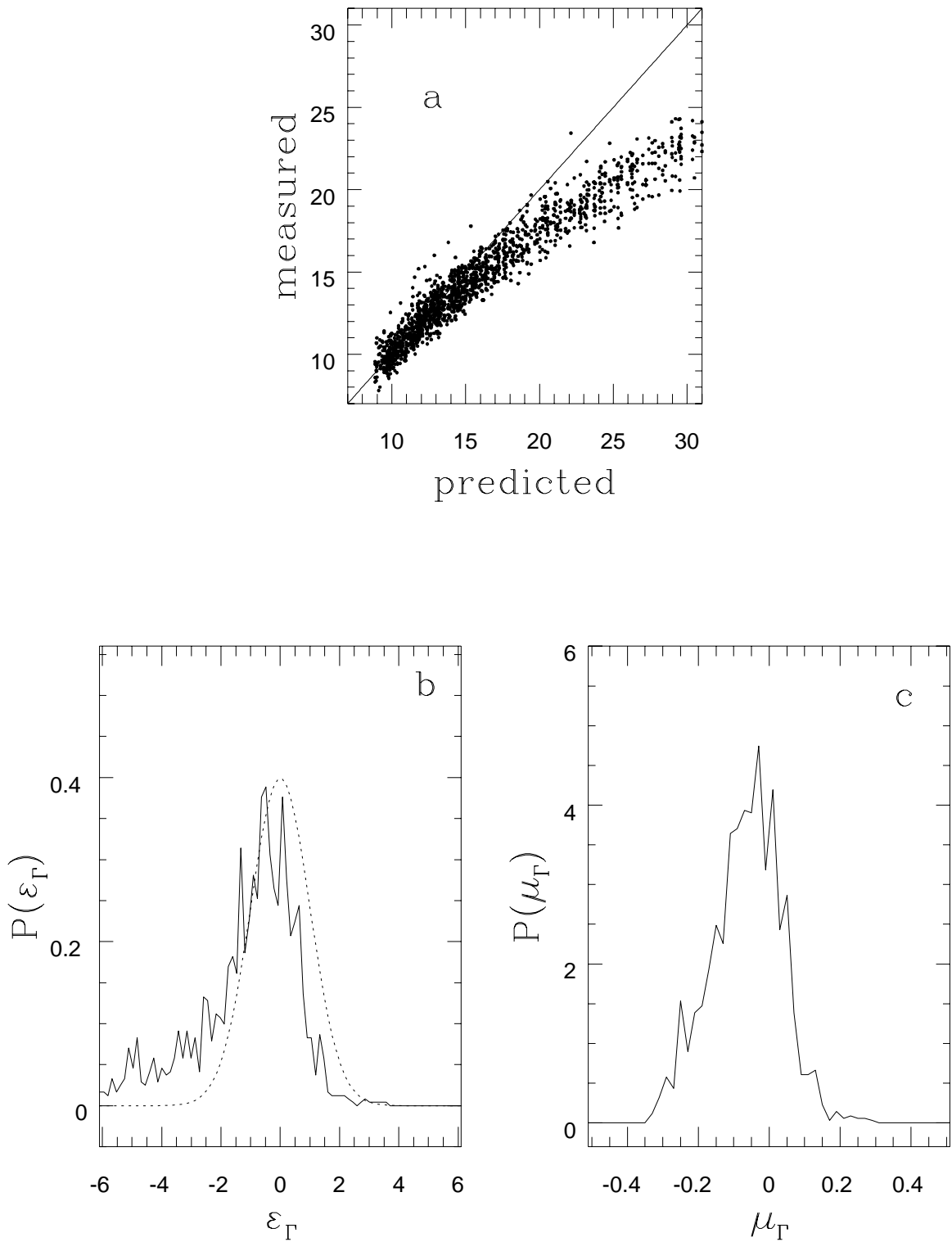}
\caption{ Measured versus predicted event rate 
$\Gamma$. The velocity distribution is assumed to be isotropic,
Maxwellian, and independent of position.  We use $v_\infty=212\kms$ as
determined from the halo as a whole.  Model I (the triaxial model) is
used for the mass distribution.  (a) Scatter plot of measured versus
predicted total event rates for the 1728 experiments.  (b)
$P(\varepsilon_\Gamma)$ as a function of $\varepsilon_\Gamma$. (c)
$P(\mu_\Gamma)$ as a function of $\mu_\Gamma$.}
\label{fig-gamma-triax}
\end{figure}

\begin{figure}
\plotone{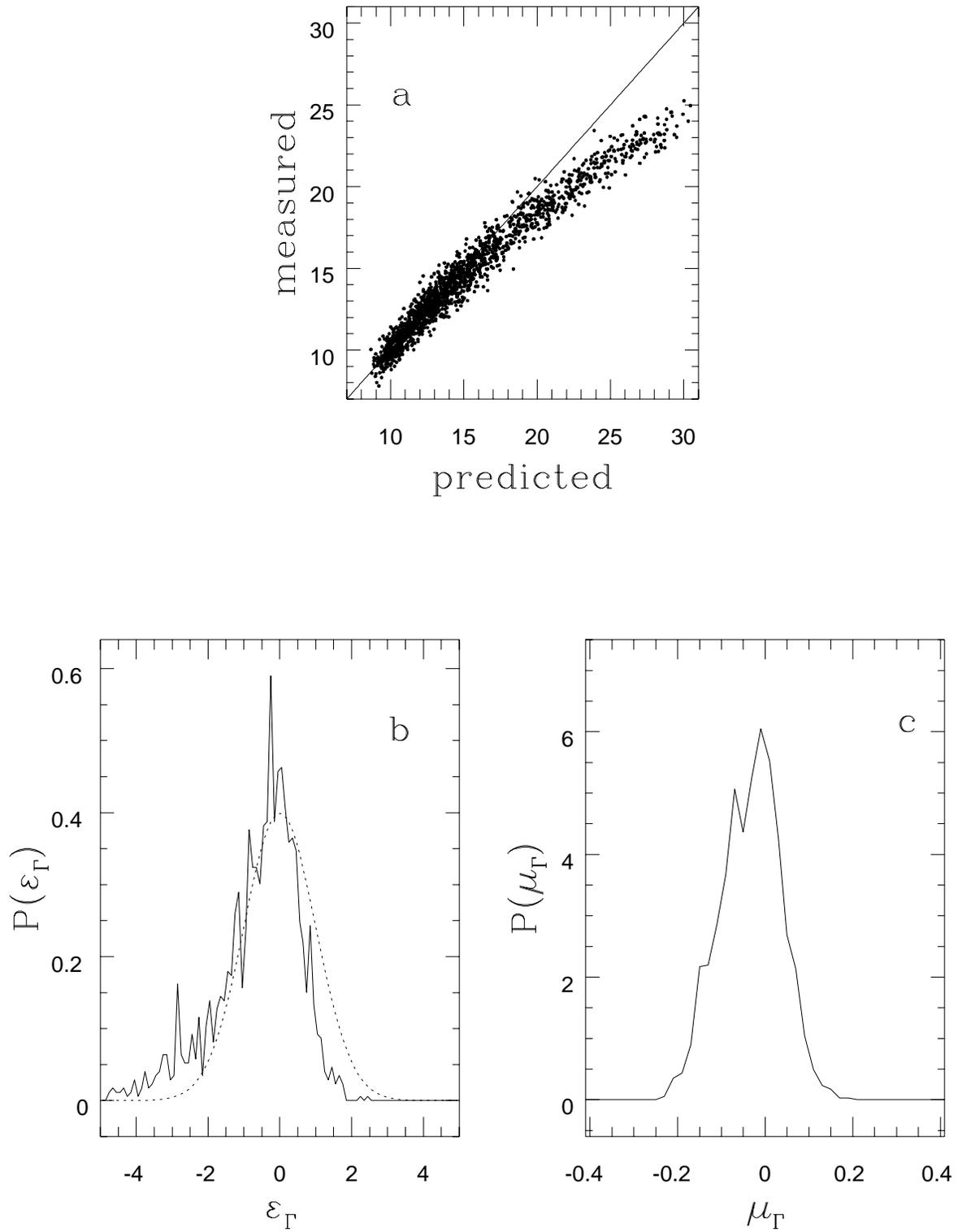}
\caption{ Same as Figure 7 but this time we
choose the rms velocity separately for each lensing tube.}
\label{fig-gamm-improve}
\end{figure}

\begin{figure}
\plotone{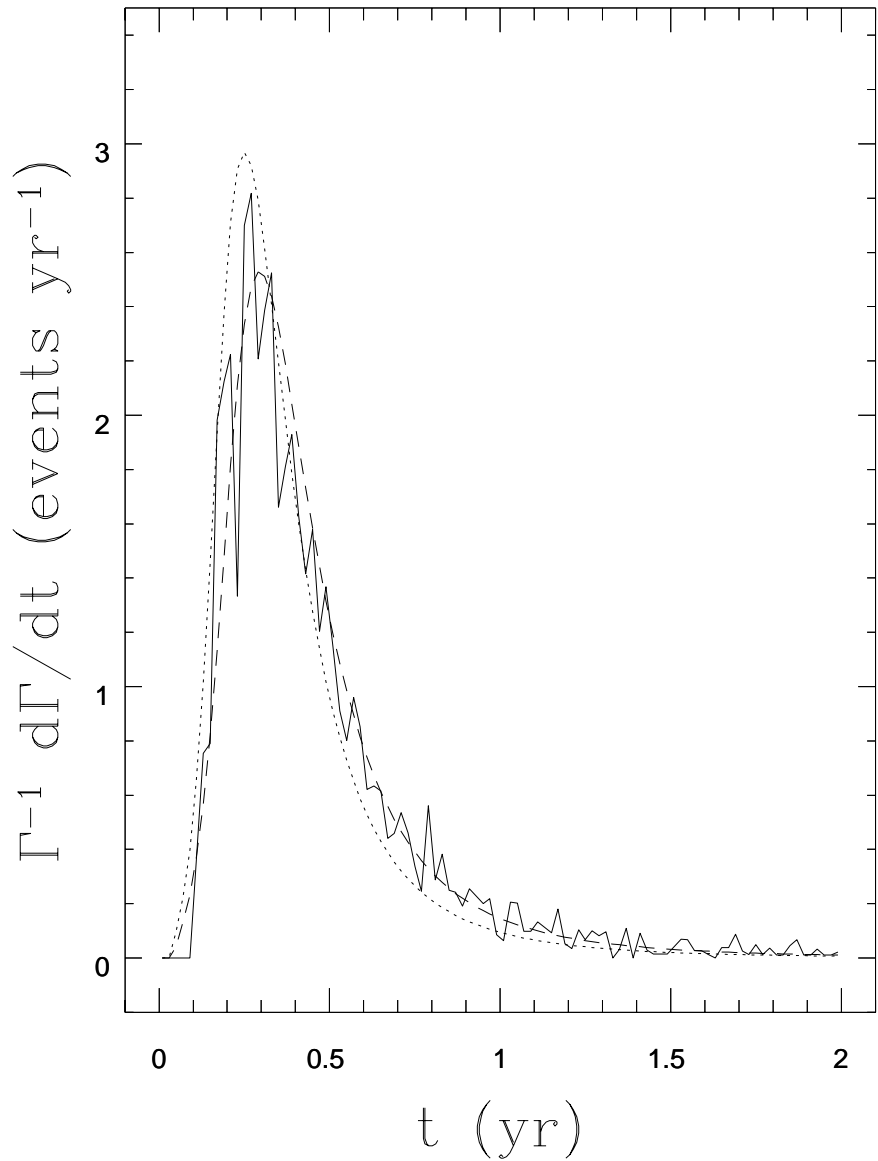}
\caption{ Event rate as a function of event
duration, $d\Gamma/d\hat t$ for one of the experiments.  The solid
line gives the measured $d\Gamma/d\hat t$.  The dotted line is the
model prediction with $v_\infty=212\kms$ (from the velocity dispersion
of all of the particles in the halo).  The dashed line is the
model prediction with $v_\infty=180\kms$ (from the velocity dispersion
in this particular lensing tube).}
\label{fig-dgamdt-anios}
\end{figure}

\begin{figure}
\plotone{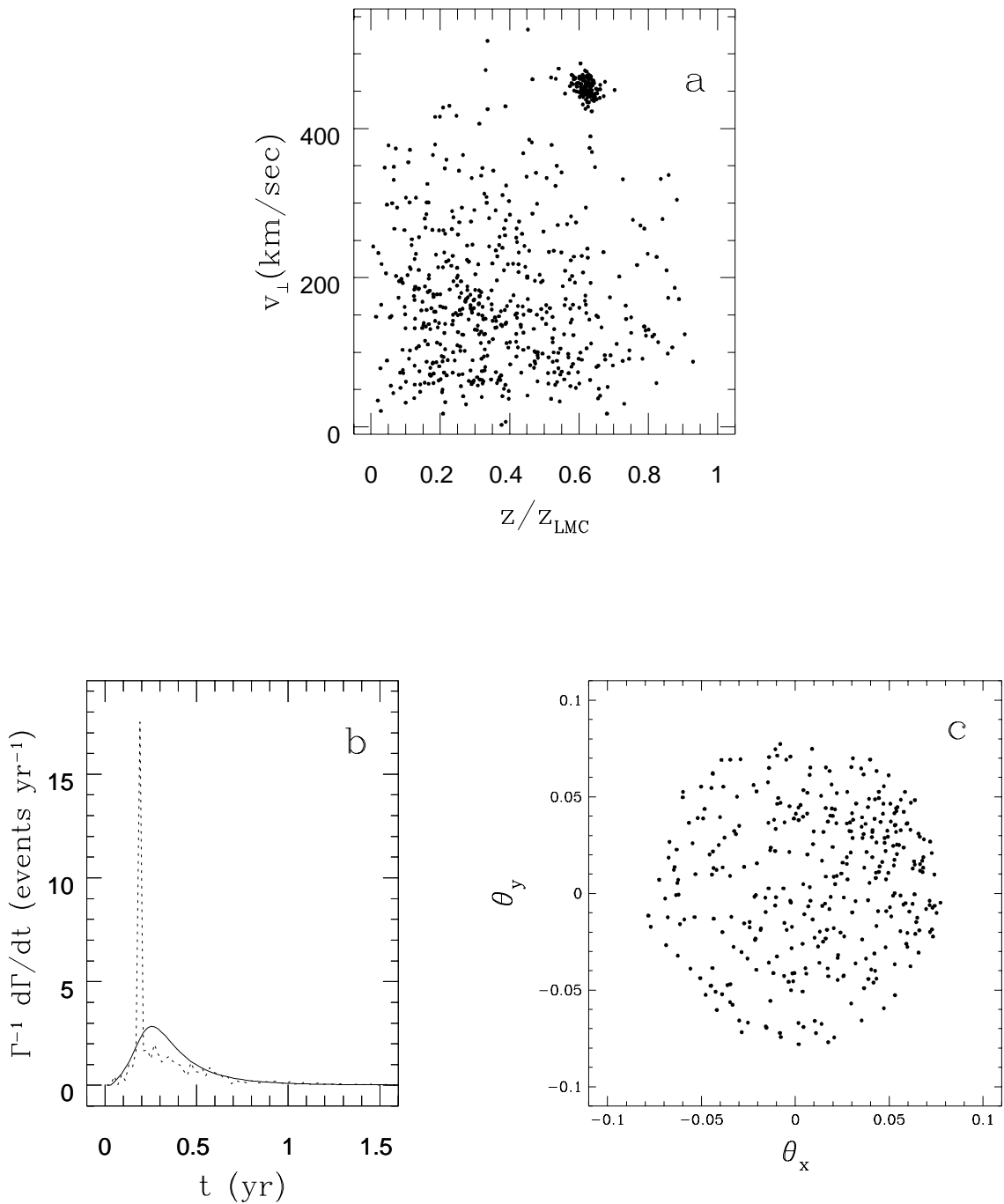}
\caption{ (a) Phase space distribution of particles
where there is a coherent clump in the lensing tube.  In this experiment the
deviation between measured and observed optical depth is over
$3\sigma_\tau$. (a) Distribution of particles in the lensing tube as a
function of $v_\vp$ and $z'$.  (b) $d\Gamma/d\hat t$ for this
experiment.  The clump would show up as an excess of $70\,{\rm day}$
events.  (c) Map of the sky in the region of this hypothetical LMC.
The plot gives the angular distribution (measured in radians) of all
objects in the foreground of the LMC.}
\label{fig-vp-vz}
\end{figure}

\begin{figure}
\plottwo{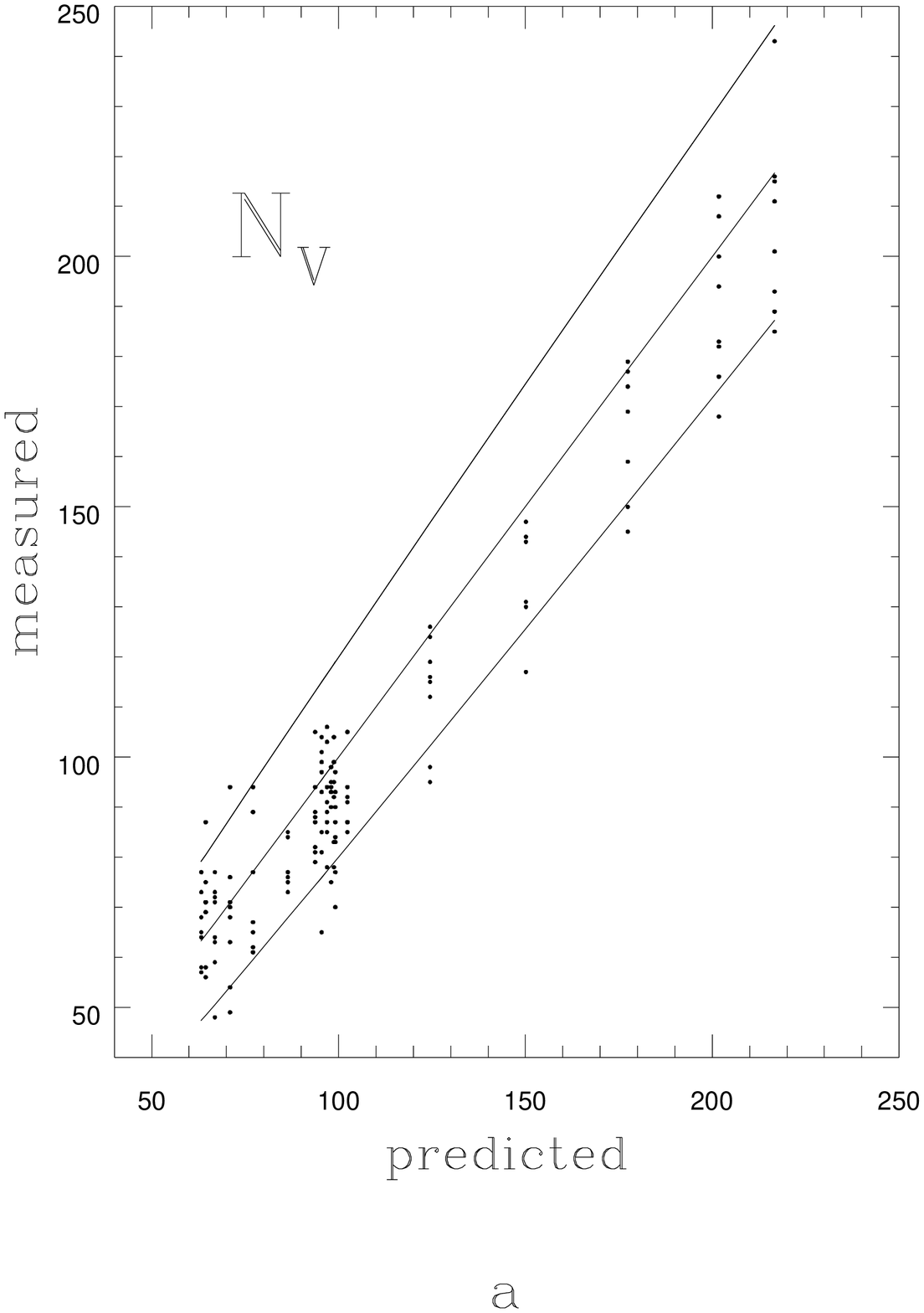}{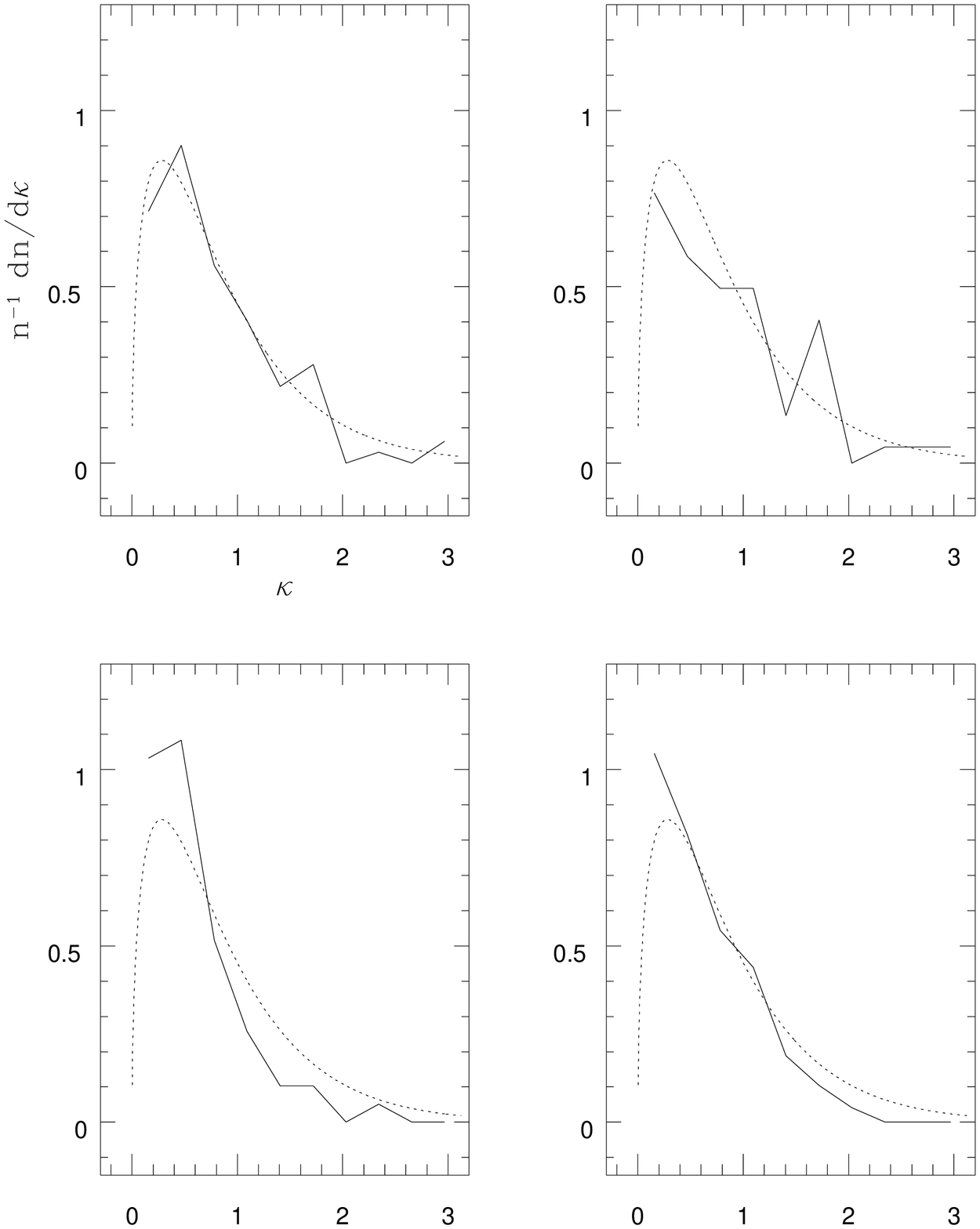}
\caption{(a) Results for $100$ dark matter search experiments.  
A comparison is made of the total number of particles measured in a
volume $V$ centered on a given observer with the number predicted by the
model.  The lower and upper solid lines give $2\sigma$ error bars.
(b) Normalized energy spectra for 4 of the observers.  The dotted
lines are model predictions assuming a Maxwellian distribution for the
velocities. }
\label{fig-wimp}
\end{figure}

Traditionally dark matter search experiments such as MACHO have
assumed some idealized model for the dark matter DF (e.g.,
Eqs.\,(1-3)) leading to potentially large systematic errors.  These
errors can be estimated by considering alternative models for the halo
such as those discussed in Section 2.  Our goal has been to test this
programme by performing microlensing experiments in an N-body
realization of a CDM galaxy and comparing the results with predictions
made assuming a variety of halo models.  Our conclusions are as
follows:

\begin{itemize}

\item The success of a gravitational microlensing experiment
depends crucially on the quality of the model chosen for the
distribution function of the Machos.  The galaxy in our simulated
experiment is highly asymmetric.  By accurately modeling the shape and
density profile of the galaxy we can achieve excellent agreement
between measured and predicted quantities such as the optical depth
and event rate.  If instead, a spherical model is used, the agreement
is rather poor, the typical systematic errors in the determination
of $\tau$ and $\Gamma$ being $30-50\%$.

\item  A wide range of fitting formulae will adequately describe
the density profile of the halo in the region between the observer and
the LMC.  In this work, we use the NFW profile though a cored
isothermal sphere, fit to the measured density profile in this region,
will work just as well.  An accurate global fit to the density profile
is required only if one plans to use observations well outside this
region.  This suggests that, for actual microlensing experiments, one
should use a model density profile based on observations of the Galaxy
between $8.5$ and $50\kpc$.  However most of these observations are of
material in the Galactic plane whereas the LMC is at high Galactic
latitude.  Systematic errors, such as those seen in
Figures~\ref{fig-gamma-triax} and \ref{fig-gamm-improve}, will arise
if the halo is far from spherical and the model density profile does
not provide a good global fit even if we have correctly modeled the 
shape of the halo.

\item  The distribution of Machos in velocity space affects the event
rate and event duration both of which are used to estimate the mass and
density of the Machos.  The triaxial shape of our N-body galaxy is
supported by velocity space anisotropy.  By properly taking this into
account one can improve the agreement between measurements and
predictions.

\item Substructure in the distribution of Machos can affect the outcome
of a microlensing experiment.  In particular, a clump of Machos
passing between us and the LMC can significantly bias the results.
Though this situation arises in only $\sim 1\%$ of our experiments, it
can lead to errors as large as $50\%$ in estimates of the optical
depth and event rate.

\end{itemize}

One should not take the quantitative results in this work too
literally.  First, the simulation was done in the context of a
standard cold dark matter universe which may not be applicable if
Machos are a significant fraction of the dark matter.  A baryon
dominated universe might be more appropriate.  Of course, these models
tend to have more power on small scales suggesting even more halo
substructure.  Second, our galaxy was created in a pure
dissipationless simulation so that there is no gaseous or stellar
disk.  Disk formation will affect the both shape of the halo and the
density profile.  For example, the presence of a disk will probably
drive our prolate halo toward a more triaxial shape (Katz \& Gunn
1991; Dubinski 1994) which might improve the performance of the
spherical model.  Nevertheless, it is probably optimistic to think
that we know the shape of the halo well enough to be able to use
Gaussian or Poisson statistics for a microlensing experiment.
Finally, our simulation, with particles of mass $M_{\rm P}=2.7\times
10^6 M_\odot$, is not able to resolve all of the substructure relevant
to a microlensing experiment.

Despite these caveats we believe our conclusions to be at least
qualitatively correct.  It will be interesting to perform microlensing
experiments in other simulated galaxies and in particular, simulations
that include gas.  In a simulated spiral galaxy, for example, one can
place hypothetical observers in the disk.  These observers can then
``measure'' the rotation curve of the galaxy and use this to model the
density profile adding another layer of realism to the exercise.
Along these lines, it would be interesting to choose lines of sight in
the simulation toward actual satellite galaxies to see if tidal debris
of the type discussed by Zhao (1997) and Zaritsky \& Lin (1997) can
indeed affect microlensing experiments.

\bigskip\bigskip

\centerline{\bf ACKNOWLEDGEMENTS}

\bigskip

It is a pleasure to thank S. Columbi, J. Dursi, G. Holder, and
S. Tremaine for useful discussions.  This work was supported in part
by a grant from the Natural Sciences and Engineering Research Council
of Canada.  LMW acknowledges the hospitality of the Canadian Institute
for Theoretical Astrophysics during a recent visit when much of this
work was completed.  JD acknowledges a supercomputing grant at the
Pittsburgh Supercomputing Center where the N-body simulations were
run.

\newpage

\appendix{\bf APPENDIX}

We begin by deriving an expression for the standard deviation
$\sigma_\tau$.  The expectation value of $\tau$ is given by
Eq.\,\EC{tau}.  In our simulated experiment, we count particles in the
conical line-of-sight volume $\Omega$.  Let $k$ be the region in
$\Omega$ such that $z_k\le z'\le z_k + \Delta_k$.  The probability of
finding a particle in $k$ is ${\cal P}_k\Delta_k =
\pi\left (\theta_{\rm LMC} z_k\right )^2\Delta_k\rho(z_k)/M_{\rm P}$.
(We assume $\Delta_k$ is small enough so that the probability of finding
two particles in $k$ is negligible.)  A particle in $k$ contributes an
amount $T_k = R^2_{\rm E}(z_k)/\left (\theta_{\rm LMC} z_k\right )^2$
to $\tau$.  We can therefore write

\be{tau2}
\langle\tau\rangle = \sum_k T_k \pkdk
\ee

\noindent The characteristic function for $\tau$ is

\be{cf}
\varphi_\tau(s) = \prod_k
\left ( 1-\pkdk +\pkdk e^{isT_k}\right )~.
\ee

\noindent Moments of $\tau$ are found by differentiating $\varphi_\tau$
with respect to $s$ and setting $s=0$.  For example
$\langle\tau\rangle = \varphi_\tau'(0)/i$ in agreement with Eq.\,\EC{tau}.
A straightforward calculation gives

\bea{secondm}
\langle\tau^2\rangle & = & \frac{\varphi_\tau''(0)}{i^2} \\
                     & = & \sum_k T_k^2\pkdk + 
                           \sum_k\sum_{j\ne k} T_k\pkdk T_j\pjdj~.
\eea

\noindent $\sigma_\tau$ is given by the expression

\bea{stdv}
\sigma_\tau & =      & \left [\langle\left (\tau - 
		       \langle\tau\rangle\right )^2
	               \right ]^{1/2} \\
            & \simeq & \left [\int dz' T^2(z'){\cal P}(z')
                       \right ]^{1/2}
\eea

\noindent where in the last line we let $\Delta_k\to 0$.

We next calculate the standard deviation $\sigma_\Gamma$
for the total event rate
$\Gamma$.  For simplicity, we assume that the velocity distribution of
the Machos is isotropic and Maxwellian.  We can write $\Gamma$ as the
following double integral:

\be{gamma2}
\Gamma = \frac{4}{M}\int R_{\rm E}(z')\rho(z')dz'\int
\frac{v_\vp^2 dv_\vp}{v_\infty^2}\exp{\left (-\frac{v_\vp^2}{v_\infty^2}
\right )}
\ee

\noindent In this case $\pkdk=\pi R^2_{\rm tube}(z_k)
\Delta_k\rho(z_k)/M_{\rm P}$ where we are using the volume
$\tilde\Omega$ rather that $\Omega$ (cf. Section 3.2).  Similarly
the probability that a particle will have $v_\vp^2$ between
$v_\infty^2 u_n$ and $v_\infty^2\left ( u_n + \Delta_n\right )$ is
$\vndn\equiv \exp{(-u_n)}\Delta_n$.  The characteristic function for
$\Gamma$ is

\be{cfgamma}
\varphi_\Gamma(s) = \prod_n\prod_k\left (
1 -\pkdk\vndn + \pkdk\vndn e^{iG_{nk}s}\right )
\ee

\noindent where $G_{kn}\equiv 2v_\infty u_m^{1/2}R_{\rm E}(z_k)/
\pi R^2_{\rm tube}$
The expectation value of $\Gamma$ is 

\be{expgamma}
\langle\Gamma\rangle =\frac{\varphi_\Gamma'(0)}{i} = 
\sum_{kn}\pkdk\vndn G_{kn}
\ee

\noindent in agreement with Eq.\,\EC{gamma2}.  The standard deviation of
$\Gamma$ is

\bea{stdvgamma}
\sigma_\Gamma  & = & \left [4 v_\infty^2\int dz' 
			\left (
			\frac{R_{\rm E}(z')}{R^2_{\rm tube}}\right )^2
			\frac{\rho(z')}{M_{\rm P}}\right ]^{1/2} \\
	       & = & 2 v_\infty\left ( \frac{M_{\rm P}}{M_{\rm tube}}
			\right )^{1/2}\left [
			\int dz'\frac{\rho(z')}{M_{\rm P}}\right ]^{1/2}
\eea

\noindent where $M_{\rm tube}/M_\odot=
\left (R_{\rm LMC}/9.65\times 10^{-8}\right )^2\left (L/50\kpc\right )$.
Had we used $\Omega$ rather than $\tilde\Omega$, we would have found
a logarithmic divergence.

\newpage

\centerline{\bf REFERENCES}

\bigskip\bigskip

\noindent Alcock, C. {\it et al.} 1993, Nature, 365, 621

\noindent Alcock, C. {\it et al.} 1995, ApJ, 449, 28

\noindent Alcock, C. {\it et al.} 1996, astro-ph/9606165

\noindent Alcock, C. {\it et al.} 1997a, astro-ph/9707310

\noindent Alcock, C. {\it et al.} 1997b, astro-ph/9708190

\noindent Aubourg, E. {\it et al.} 1993, Nature, 365, 623

\noindent Bertschinger, E. 1985, ApJS 58, 39

\noindent Dougherty, B. L. 1995, in Particle and Nuclear
Astrophysics and Cosmology in the Next Millennium, eds. E. W. Kolb \&
R. D. Peccei: World Scientific, Singapore

\noindent Dubinski, J. 1996, New Astronomy, 1, 133

\noindent Dubinski, J. 1994, ApJ, 431, 617

\noindent Dubinski, J. \& Carlberg, R. G. 1991, ApJ, 378, 496

\noindent Evans, N. W. 1996, MNRAS, 260,191

\noindent Evans, N.W.,\& Jijina, J. 1994, MNRAS, 267, L21

\noindent Evrard, A. E., Summers, F. J., \& Davis, M. 1994, ApJ, 422, 11

\noindent Fillmore, J. A. \& Goldreich, P. 1984, ApJ 281, 1

\noindent Frieman, J. A. \& Scoccimarro, R. 1994, ApJ, 431, L23

\noindent Gates, E., Gyuk, G.,\& Turner, M. S. 1995, ApJ, 449, L123

\noindent Gilmore, G., Wyse, R. F. G.,\& Kuijken, K. 1989, ARAA, 27, 555

\noindent Griest, K. 1991, ApJ, 366, 412

\noindent Hartwick, F. D. A. 1996 in Formation of the Galactic Halo\ldots Inside
and Out, eds. H. Morrison \& A. Sarajedini 

\noindent Holder, G. P. \& Widrow, L. M. 1996, ApJ, 473, 828

\noindent Ibata, R. A., Gilmore, G., \& Irwin, M. J. 1995, MNRAS, 277, 781

\noindent Katz, N. \& Gunn, J. E. 1991, ApJ, 377, 365

\noindent Maoz, E. 1994, ApJ, 428, 454

\noindent Majewski, S. R., Munn, J. A., \& Hawley, S. L. 1994, ApJ, 427, L37

\noindent Merrifield, M. R. \& Olling, M., 1997, private communication

\noindent Metcalf, R. B \& Silk, J. 1996, ApJ, 464, 218

\noindent Navarro, J. F., Frenk, C. S., \& White, S. D. M. 1996, ApJ, 462, 563

\noindent Paczynski, B. 1986, ApJ, 304, 1

\noindent Rosenberg, L. J. 1995, in Particle and Nuclear
Astrophysics and Cosmology in the Next Millennium, eds. E. W. Kolb \&
R. D. Peccei: World Scientific, Singapore

\noindent Sackett, P. D. \& Gould, A. 1994, ApJ, 419, 648

\noindent Sikivie, P., Tkachev, I. I., \& Wang, Y. 1996,
Phys. Rev. Lett., 75, 2911

\noindent van der Marel, R. P. 1991, MNRAS, 248, 515

\noindent Warren, M. S., Quinn, P. J., Salmon, J. K., \& Zurek,
W. H. 1992, ApJ, 399, 405 

\noindent Wasserman, I. \& Salpeter, E. E. 1994, ApJ, 433, 670

\noindent Zaritsky, D. \& Lin, D.\,N.\,C. 1997, astro-ph/9709055

\noindent Zhao, H.-S. 1996, astro-ph/9606166

\noindent Zhao, H.-S. 1997, astro-ph/9703097

\newpage

\bigskip

\bigskip

\bigskip

\bigskip

\bigskip

\bigskip

\bigskip

\bigskip

\bigskip

\end{document}